Submitted in Partial Fulfilment of the Requirements

for the Degree of

Doctor of Philosophy

Presented to the

# Swiss School of Management

# The behavioral intention to adopt proptech services in Vietnam real estate market

by

**Le Tung Bach**

Student ID: Ssm/Vietnam/phd/2019/00061

Supervisor: Professor Dr. Premkumar Rajagopal

September, 2022

# COPYRIGHT STATEMENT

Le Tung Bach

Date: Sep. 2022

# SUPERVISOR DECLARATION STATEMENT

I have reviewed the content and presentation style of this thesis and declare it is free of plagiarism and has sufficient grammatical clarity to be examined. To the best of my knowledge and belief, the research and writing are those of the candidate except as acknowledged in the Author Attribution Statement. I confirm that the investigations were conducted in accordance with the ethics policies and integrity standards of the Swiss School of Management and that the research data are presented honestly and without prejudice.

Sep. 2022

. . . . . . . . . . . . . . . . . . . . . . . . . . . . . . . . . . . . . . . . . . . . . . . .

Date Professor Dr. Premkumar Rajagopal

“To see a World in a Grain of Sand

And a Heaven in a Wild Flower,

Hold Infinity in the palm of your hand

And Eternity in an hour.”

— William Blake, Auguries of Innocence

*To my Dear Family,*

# ABSTRACT

Purpose

One of the main stages for achieving success is the adoption of new technology by its users. Several studies show that Property technology (so-called Proptech) is advantageous for real estate stakeholders. Hence, the purpose of this paper is to investigate the users' engagement behavior to adopt Proptech in the Vietnamese real estate market.

Design/methodology/approach

To that end, a purposive sample of 142 participants was recruited to complete an online quantitative approach-based survey. The survey consisted of a modified and previously validated measure of acceptance based on the extended demographic version of the unified theory of acceptance and use of technology (UTAUT), as well as usage scale items.

Findings

The empirical findings confirm that participants were generally accepting of PropTech in the Vietnamese real estate market. The highest mean values were associated with the subscales of effort expectancy and performance expectancy, while we can easily identify the lowest mean value in the social influence subscale. The usage of PropTech was slightly more concerned with the gathering of information on properties and markets than transactions or portfolio management.

Practical implications

This study provides an in-depth understanding of Proptech for firms' managers and marketers. Online social interactions might be either harmful or fruitful for firms depending on the type of interaction and engagement behavior. This is especially true of property portals and social media forums that would help investors to connect, communicate, and learn. Findings can help users to improve their strategies for digital marketing.

Originality/value

This is likely to be the first study that examines the moderating effect of users' acceptance and use of technology relying on users' engagement behavior. By providing robust findings by addressing issues like omitted variables and endogeneity, the findings of this study are promising for developing new hypotheses and theoretical models in the context of the Vietnamese real estate market.

# ACKNOWLEDGMENTS

I have been able to complete this research program with the support and active cooperation of concerned bodies and authorities and several persons. I owe my debt and would like to express deep feelings of gratitude to my teacher and thesis supervisor, Professor Dr. Premkumar Rajagopal, President at Malaysia University of Science and Technology Kedah, Malaysia for his contribution in providing insight and expertise that significantly assisted the research. His guidance at different stages of this research program is very helpful and valuable. It would have been quite impossible to carry on the research work and make it into the final shape of a thesis without his able guidance and sympathetic encouragement. His affection for me is fondly remembered.

I am intensely indebted to my associate partner Mr. Do Thanh Tung for his very useful guidance and counseling in overcoming various bottlenecks during the study and also for his comments on the draft of my thesis. His cooperation and assistance in the overall presentation and all his precautionary measures made the methodology of this thesis comprehensive. I appreciate his continuous encouragement.

My gratitude is also due to Dr. Duong Tuan Anh, Foreign Trade University for his extraordinary support of the whole process of this research project. I immensely benefited from his continuous assistance in finalizing the questionnaire, research protocol, pre-testing the cases, developing skills in interpretive approach, selecting the sample as well as his sincere guidance throughout the various stages of this study.

I am also grateful to my former employer for her support for the research work. Especially, her suggestion for incorporating additional insights for the enrichment of data was a significant contribution to this thesis. I am truly blessed for her invaluable contextual insights.

I am extremely indebted to Dr. Dao Toan, Vietnam Polytechnic University for his valuable comments on my research project and who guided and assisted me in the preparation and development of the questionnaire for this study. I would also like to acknowledge Professor Viet Le's help. His assessment, comments, and suggestions for strengthening the methodological approach were significant to improve the benchmark of this thesis.

I would also like to thank Dr. Kieu Viet a bilingual scholar, who extended his support as a panel member for reviewing the translated Vietnamese version of the questionnaire for its authenticity.

My thanks are due to the managers of Real estate brokerage firms in Hanoi - Vietnam who furnished all the relevant information and data for the study without which this research could not have been possible. I express my gratitude to all those officers and employees of TIEN KHA Real estate Joint stock company whose significant contributions were evident in this thesis in accumulating data as well as boosting my spirit when it was needed most.

I gratefully acknowledge the financial support received from the Doctoral Fellowship from the Research Department of TIEN KHA Real estate Joint-stock company. I would also like to thank Mr. Kha President and CEO of the company for granting me study leave for the whole period of my Ph.D. studies.

Further, my sincere thank goes to Mr. Khang Le and Ms. Helen Pham for their sincere efforts in proofreading this thesis at various stages including the final draft.

Finally, my heartiest thanks go to my family, my wife THUY THAN, my son Leo and daughter Lea whose proximity, love, and affection provided me joy and relaxation though they sacrificed a lot of my company. Especially, I would like to thank my wife who appreciated my efforts and always provided cheerful encouragement during the period of this study. I have spent three and half years in Dong Anh district which was a long as well as hard time for my parents, Mr. Le Hung and Ms. Le Hau, whose everyday prayers made it possible to complete this thesis.

# TABLE OF CONTENTS

# LIST OF FIGURES

# LIST OF GRAPHS & TABLES

# Chapter I: INTRODUCTION

In October 2021, International Monetary Fund (IMF) forecasts the Gross Domestic Product (GDP) of Vietnam is expected to skyrocket exponentially to an astounding USD 630.5 billion by 2026. Similarly, projections indicate that there will be 120 million citizens living in Vietnam by the year 2050, an increase of population from more than 97.33 million people living in Vietnam in the year 2020. Thus, the country is currently going through a period of fast social and demographic transformation, which subsequently bring up a clear demand for residential products in the years to come.

However, despite its growth, Vietnamese users (organizations and professionals/ individuals) have been slow to adopt digital technologies. In the Vietnamese real estate industry, the transaction still requires a lot of paper works, documents, and face-to-face/phone meetings to relay information compared to most developed countries which would be digitally processed. There is little information about the use of Proptech in Vietnam to assist in the acquisition and management of real estate portfolios. This thesis research will study in details possible issues and variables related.

This section of the thesis serves as an introduction which covers the research problem, research objectives and questions, scope of the study/delimitation, conceptual framework, research methods and analyses and major areas of contributions.

This introductory chapter is designed to:

- Review the background of the research and analyze Proptech concept from a variety of viewpoints and apply them to current Vietnam’s real estate market scene (Section 1.1);
- Analyze the Rationale of the research study in Section 1.2.;
- Define the study’s concepts in Section 1.3.;
- Short summary of the hypotheses in Section 1.4.;
- Define the statement of the problem in Section 1.5. and research objectives and questions in Section 1.6.;

- Suggest the research methods and analysis (Section 1.7.);
- Specify the scope and the limitation of the study in Section 1.8.;
- Identify significance of the study in Sections 1.9.;

The growing number of people in Vietnam who have access to the internet and a rise of young middle-class with disposable income have contributed to a growth in the demand for residential real estate and the proportion of sales that take place through online channels.

Because of this sudden growth in individual wealth, many Vietnamese individuals are now in a position to own real estate, which has contributed to both an increase in the number of new developments and an increase in property prices.

Due to the fact that urbanization has generated a continual need for housing in large urban areas, the focus of the residential market has recently changed from the high-end to the mid-value portions of the market. As a result of the country's expanding economy and the passage of legislation that makes it less difficult for non-citizens to buy property there, the nation is now largely regarded as the premier location for investing in luxury real estate.

On the other hand, according to the findings of a survey conducted by Propzy insight, a firm that specializes in real estate brokerage, the typical age of a person purchasing their first home is 37 years old, and it is anticipated that this average age will decrease in the near future.

## 1.1. Background of the study

Vietnam has recently witnessed an incredible growth on numerous industries comparing to other Southeast Asian economies and particularly construction and real estate are among the most potential sectors to rise. And, it is also expected to be in continuous growth in the years to come. The optimistic anticipation is mostly due to the country’s sustained fast urbanization coming along with robust economic growth and the implementation of multiple mega-projects in key cities such as Hanoi and Ho Chi Minh City.

One of the most rapid urbanization speed over the globe currently taking place may be seen in Vietnam. Over the past thirty years, the city's population has increased by a factor of two. Since 2010, it has been rising at roughly 3% per year, placing Vietnam's urbanization rate above the yearly average for Southeast Asian countries (2.5%), and extremely similar to the rate China experiences (3.1%) (source: World Bank)

The demand for residential products has increased as a result of the widespread growth of cities, which has encouraged an increase in the number of people moving from smaller towns to larger cities. The population of Vietnam was estimated to reach 120 million by the year 2050, up from the current estimate of 97.33 million in 2021. Statistics provided by the World Bank indicate that Vietnam's urban population was 31.1% in the year 2011, and it is projected to reach 37.1% in the year 2021.

The United Nations' most recent forecasts indicate that by the year 2039, fifty percent of Vietnam's population would live in urban areas, and that percentage will rise to sixty percent by the year 2050.

This transition from a rural to an urban society is intimately connected to the socioeconomic reforms that were initiated in the middle of the 1980s. These changes progressively liberalized the economy and eased the grip that the state had on the movements and activities of the populace. These reforms, which were given the name Doi Moi (which roughly translates to "new change"), eased restrictions on the movement of people from rural to urban areas and made it possible for occupations to transition away from agriculture.

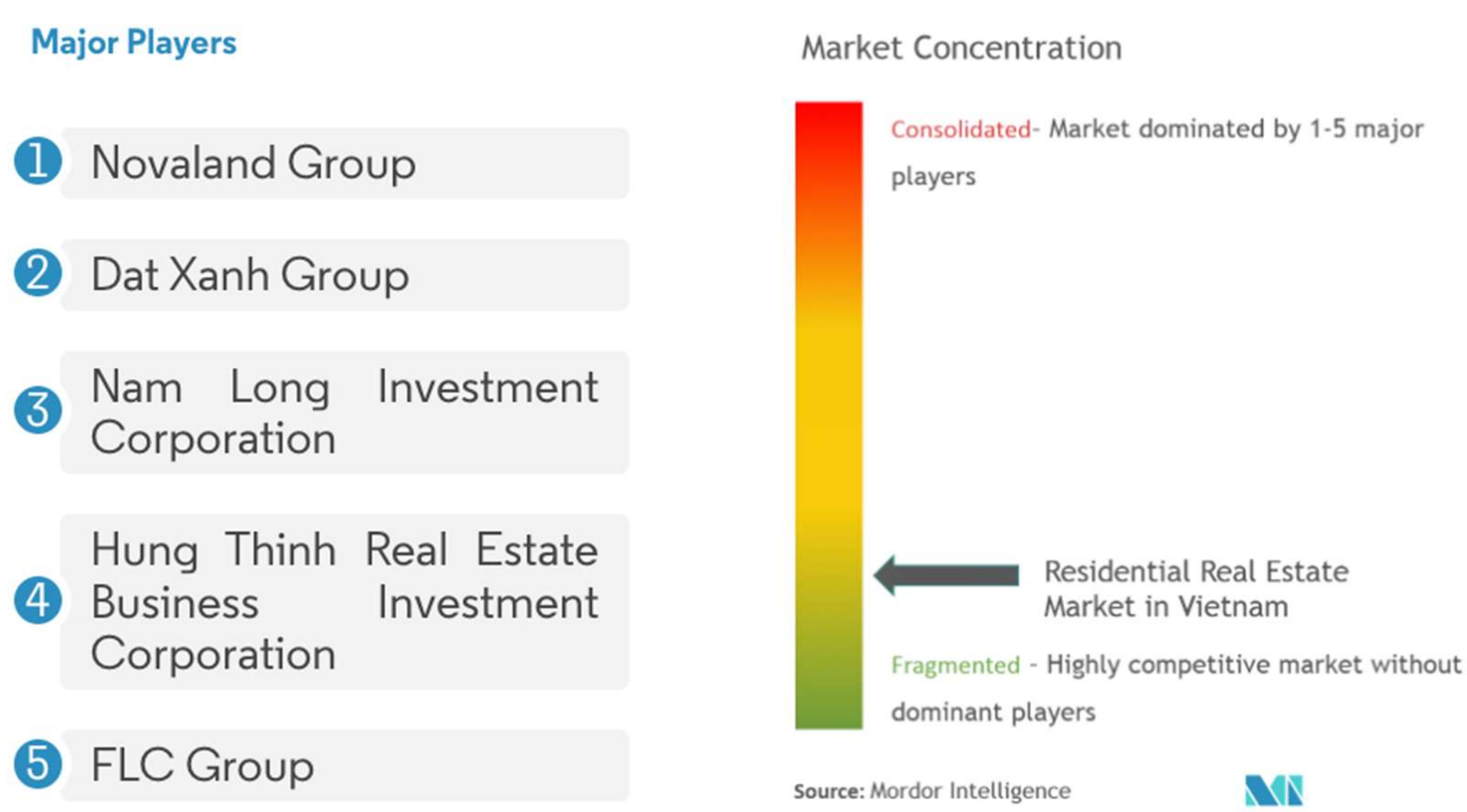

*Figure 1.1.b – Real estate market concentration*

(Source: Mordor Intelligence)

As above figure shown, Mordor Intelligence's survey reports 5 biggest key companies in Vietnam real estate market are Dat Xanh, FLC, Novaland, Hung Thinh, and Nam Long.

**Covid-19 outbreak**

The epidemic caused by COVID-19 had a negative effect on the housing markets in Vietnam. As a result of the pandemic, Savills Hanoi (a real estate company) reported that Hanoi only introduced 4,800 new homes in the first quarter of 2020, which was the lowest number in the past five years. Even before the COVID-19 breakout, the real estate market had already begun to see some degree of stagnation. The situation has been increasingly worse over the past two years. The pandemic put a significant strain on businesses, making it difficult for them to continue operating normally. During this time, the government came to the conclusion that increasing public investment was an essential component to bolster economic expansion. Therefore, an improved infrastructure system might be considered as positively encouraging news for the real estate market for continued expansion in the coming years.

The pandemic caused by COVID-19 fortifies and technologically recalibrates the concept of home. Responses that involve staying at home are predicated on the assumption that the public realm is both dangerous and unpredictable, and that the house is both a safe haven and a fortress. Because of the data that digital technologies generate, we do not need to leave the comfort of our home. (Maalsen, 2020).

## 1.2. Rationale

**a. Vietnam Proptech ecosystem**

Upon on the definition of Forbes, Proptech involves all the companies which leverages technology to enhance the structure we buy, rent, sell, design, construct, and manage residential and commercial property. Despite the fact that it is still a relatively new industry, the global Proptech startup community has already raised over $43 billion in capital since 2012. (Deloitte, 2021).

According to Marco Breu, managing partner of McKinsey & Company Vietnam, Vietnam has just recently reached a phase that may be described as a "golden age". The Ministry of Planning and Investment of Vietnam has made the public announcement in September 2019 about the draft of Vietnam's national strategy for the Fourth Industrial Revolution. Since then, the nation has progressed to be leading country the region in terms of technological advancement across a variety of business sectors.

There are now 56 real estate technology companies operating in Vietnam, according to FinREI Investments JSC, a Proptech startup that focuses on the real estate industry. These companies offer a variety of services detailed in the following figure:

According to JLL Vietnam, over 80% of the companies in the industry are either international startups or startups sponsored by investors from other countries.

Overall, the analysis shows that no concerted effort has been made in the literature to investigate the behavioral intention to adopt Proptech services in the Vietnamese real estate market. There is a lack of agreement between users on the impacts of Proptech which, consequently displayed a little of studies on related topics. A lot of discovery data are conjectural and need to be confirmed in the following studies. Hence, the research problem is under the effort to be identified in the following section.

Furthermore, this finding is consistent with the findings that were previously determined. The limited number of studies that have been conducted in this and related areas have also been hindered by the fact that users disagree on the influence that Proptech services have. A good number of the findings are also speculative until they have been confirmed in subsequent investigations.

**b. The Study in Context**

From the searching of numerous academic literatures, it has been shown that there is no concrete study on the intention to adopt Proptech in Vietnam, a developing country context. Similarly, studies on innovation, realtor, technology, property management, IoT, etc. have also not focused on this topic as potential research.

Nevertheless, this setting for research is significant since the intention behavior in such developing countries could be fascinating in the context of the process of technology adoption. Therefore, the study in such a scenario is vital for real estate developers and policymakers in order to have a thorough grasp of the situation. This study focuses on factors from the perspective of a developing country, despite the fact that Proptech is a vast field of interest for international business researchers. As a direct result, the author has made the decision to include the theoretical environment for the variables of interest.

## 1.3. Definition of Concepts

The unified theory of acceptance and use of technology (UTAUT) is considered to be the most exclusive theory which incorporates aspects of both the Technology Acceptance Model (TAM), the Technology Readiness Assessment (TRA), and the Motivational Model (Davis, 1992).

The UTAUT model suggests the determinants of the behavioral intention to adopt the technology are social influence, facilitating conditions, effort expectancy, and performance expectancy.

It is possible to deduce the evolving behavioral intention based on the discussions that were offered earlier with the help of the conceptual framework.

After conducting a comprehensive examination of the relevant conceptual and theoretical literature, it became clear that there was a pressing requirement to create an alternative model in order to empirically test the building process. As a consequence of this, the core framework that was proposed earlier contributes to the development of additional models.

## 1.4. Hypotheses

From the perspective of this research, following hypotheses have been proposed:

- H1: Performance expectancy is an influence on users' intention to adopt Proptech.
- H2: Effort expectancy is an influence on users' intention to adopt Proptech.
- H3: Social influence is an influence on users' intention to adopt Proptech.
- H4: Facilitating condition influences on user's intention to adopt Proptech.

## 1.5. Statement of the problem

American real estate services firm Jones Lang LaSalle (JLL) and the real estate investment manager LaSalle Investment Management published their biennial Global Real Estate Transparency Index report in July 2020 which suggests Vietnam's improvement to regulatory reforms, embracing market data, and sustainability initiatives. In the real estate transparency index for this year, Vietnam is now ranked 56th out of 99 economies, moving up five spots from its previous position. However, the real estate market in Vietnam is not at high level of efficiency and transparency because all real estate transactions are currently based on the skills of salesmen and brokers without any assistance from automation.

(Wilson, 2018) has shown that companies will achieve the greatest possible improvement in performance if they combine the efforts of humans and machines in their operations.

As one of the most advanced technology in Proptech, the implementation trend of Artificial Intelligence and automation have delivered a significant improvement in working conditions and a great impact on the real estate, finance and banking sector. (Bughin, 2018) stated in the publication of McKinsey that these technologies “will provide multiple benefits in the form of higher productivity, GDP growth, improved business performance, and new prosperity, but they will also modify the skills that are required of human workers”.

The report by PwC (Rao A. , 2017) anticipated the contribution of artificial intelligence could be more than $15 trillion to the global economy between the years 2017 and 2030. In addition, research conducted by (Stamford, 2019) Gartner titled "2019 CIO Survey: CIOs Have Awakened to the Importance of AI" discovered that 37% of businesses have already integrated some type of artificial intelligence into their daily operations. As a result, the real estate industry is in an excellent position to take use of artificial intelligence (AI) and automation technologies to boost productivity, save costs, and reduce the number of errors.

(Ullah F. S., 2021) aims to capture the users’ perception of Real estate Online Platform and also suggests future studies related to developing countries for comparison. The risk and barrier with the technology platform also be suggested for future studies.

Especially with the coming of Covid-19, (Porter, 2019) analyzes opportunities and challenges for real estate brokers in post Covid-19 period by examining the co-emergence of Proptech and Big Data, the impact on land and housing dynamics, and the implications for planning governance and systems. The result shows that Digital technologies utilizing big data and artificial intelligence will remain on the sidelines of planning practice. As technologies mature and become more sophisticated their utility will be exploited to aid design and assessment.

The dilemma of comprehending Proptech needs to be cleared and outlined in the unique context of the Vietnamese real estate market, based on the research background that was mentioned in section 1.1 and 1.2. The central research question underpinning this thesis was:

"How do users in the Vietnamese real estate market perceive and utilize Proptech to research, acquire, and manage their holdings?"

Therefore, it might be worth looking into as part of a more integrated and all-encompassing comprehension of the real estate market in Hanoi's capital city, which is one of the two largest real estate markets in Vietnam. The research topic and primary purpose of this investigation are going to be discussed in the following section.

## 1.6. Research objectives and questions:

Following the background of the study and an overview of the research problem, there appears to be an immediate need and opportunity to review the findings of a preliminary quantitative method research study. The study's objective was to gain an understanding of how users in the Vietnamese real estate market perceive and make use of property technology (also known as Proptech) in order to conduct research, acquire holdings, and manage their properties. As a result, it has set its sights on accomplishing the following objectives, and in order to accomplish each objective, the following research question was submitted for this study:

| Specific Objectives: | Questions: |
|---|---|
| To understand how users in the Vietnamese real estate market perceive Proptech | What is users' acceptance of Proptech in the Vietnam Real estate market? |
| To find out the actual usage of Proptech in the Vietnamese real estate market | What is users' actual usage of PropTech in the Vietnamese real estate market? |

*Table 1.6.* - ***Research Objectives and questions***

This study aims to build and test a complete conceptual research framework that evaluates the adoption of Proptech based on relevant theories and the literature in order to answer the research question that was presented earlier in the introduction. The specific purpose of this research is to evaluate the behavioral intentions of Vietnamese real estate market participants about the adoption of Proptech services.

## 1.7. Research Methods and Analyses

Quantitative methods have been applied in order to successfully test the research framework. In a nutshell, primary data were collected from 145 different real estate agents in a developing nation for the purpose of quantitatively evaluating the concept (Vietnam). Only 142 cases were chosen for the study in the end, because of the removal of cases with missing data and outlier cases. The Author uses a number of different types of measurements to evaluate the proposed models which reveal that the proposed model with some modifications was better befitting than the competing model

## 1.8. Scope and Limitations of the Study

**a. Scope**

The investigation that is being proposed brings together a number of fundamental theoretical considerations to offer a clearer conceptualization of Proptech within the framework of an internationalization dyad, particularly as it pertains to the context of developing countries. A strategy along these lines is recommended to take into account: The majority of the identified factors have been investigated from a variety of viewpoints, and this study combines the findings of previous studies that are relevant to this one and concentrates on those studies. Because of this, the variables have been provided with auxiliary support in the process of conceiving the new context. In order to generalize the findings, cross-sectional data have been gathered from a developing nation.

Both the conceptual model that was offered and the competing model that was being considered as an alternative were put through a series of tests to validate the mediating effects of certain predictor variables in the model as well as their respective levels of explanatory power.

**b. Limitation**

Because this research is limited to the analyses and investigations of the intention to adopt Proptech, based on the literature review, there are certainly other issues that might have impact on the adoption of Proptech. Exogenous elements are likely to include a wide variety of macro concerns, including those that are political, legal, cultural, and environmental in nature.

Before making any conclusions from the data, it is essential to bear in mind the research's limitations, which are commonly characterized by different types of investigation. The key area of focus is centered on determining the application to a wider population. Second, this study is a cross-sectional study that evaluates the behavior of real estate stakeholders at a single point in time. This makes it possible that the results are not as applicable as those of a longitudinal study.

## 1.9. Significance of the Study

The study is important for a variety of different reasons. The most important contribution made by this study can be understood largely through both theoretical and practical lenses, specifically through managerial, marketing, and educational ramifications. These are the areas where the study has had the biggest impact.

From a theoretical perspective, this study investigated, proposed, and tested a UTAUT model, which is beyond the concept of technology acceptance, to fill a vital gap in the professional and academic literature that can help developers of PropTech and government agencies to improve their services and products for investors.

Practically, the knowledge gained from this research will provide PropTech providers and government agencies with the knowledge to improve upon existing PropTech and develop new products in this sector. In addition, the modified survey used in this chapter can be adopted for further studies, both by academics and professionals, to further study the phenomenon of PropTech in Vietnam.

All industry players will benefit from the study, as such:

- General population/potential buyers
- Real estate Developers
- Real estate Broker Agencies or independent Agents
- Government regulatory bodies

This study also helps to build up business models for:

1. Developing trust and brokerage model in Vietnam,
2. Home rental solutions,
3. Smart offices and smart homes,
4. Housing/office data analysis in the nature of the Vietnamese market,
5. Other property management applications, where users need increasing efficiencies and cost savings.

PropTech topic, which has not received a lot of attention from academics previously, is chosen as the primary research subject in this study, which makes it an important academic contribution. During this interim period, the areas of artificial intelligence and robotics, as well as blockchain and cryptocurrencies, have garnered the majority of the attention due to the significance they hold in relation to the Fourth Industrial Revolution or the growth of financial markets and information and computer technology (ICT). To the best of our knowledge, this study has garnered a limited amount of attention from other researchers. This study serves as a request for more research on the aforementioned subject. Furthermore, empirical research was used to combine and present the existing theories in this study. The results show the necessity for further research to establish a continuous Proptech service platform. Increasing the value in order to create sustainable chances for business.

From the policy maker's standpoint, Proptech and related innovation help to bring a fundamental change on controlling the real estate investment speculation. Additionally, the information gap between customers and service providers has been bridged thanks to this technology. This aspect contributes to a rise in the market's instability and impedes its efficiency. As a result, the purpose of this research is to suggest the proliferation of services that are based on PropTech as a way to rectify such information imbalance and to discover practical consequences by offering a business model that is sustainable for future PropTech companies.

## ***Chapter Synthesis***

The layout design of this thesis research is fundamentally based on the suggestion of (Perry, 1998). The next step in the research process is providing a summary of the broader perspective on internationalization that led to the primary research question. The identification of this problem is based on a conceptual framework that is supported by the theory and the literature, and it is intended for empirical testing as well as conceptual validation. In this section, both the broad outline and the organizational pattern of this thesis are broken down and addressed. The thesis is broken down into eight different chapters, each chapter is summarized as follows.

In the first chapter, "Introduction," we investigate the idea of commitment, as well as the research context and problem, the research question, the objective of the research, a framework that is based on background literature, the scope and significance of the research, a brief methodology of the study, the research context, and the expected contributions of the study.

The literature review in Chapter II concentrates on four important dimensions, each of which helps to integrate the review of the theories and overall focus on the aim to adopt Proptech.

In Chapter III, "Theoretical Background," a conceptual model with hypothesized links and a framework for conceptual understanding are developed to investigate the unfolding facts and the method by which they were developed based on the data.

In Chapter IV, the methodology of the study includes all of the important aspects of the quantitative research method that are going to be utilized in this investigation. This chapter will cover the explanation for using the quantitative method and other items related to this method.

The sample profile, the construction of the measurement model, and the testing of the proposed structural models are the three primary sections that are included in Chapter V. In the sample profile, the demographics, replies, data cleaning, and descriptive statistics of the sample are broken down in depth. Second, as part of the procedure for validating measurements, every single construct measurement has been evaluated separately and in order as part of an overall measurement model to ensure that all of the measurements are unidimensional.

In the end, proposed models as well as competing models were put through the rigors of Structural Equation Modeling (SEM) and compared to the total measurement model to validate whether or not the data and the theory are a good fit overall. Additionally, there are findings on the research's construct validity and reliability included in this particular segment of the research.

The findings and implications are presented in chapter VI with specific findings, hypotheses results. This chapter provides a synopsis of the general findings of the research. The main objective of this chapter is to build up the explanation for the question and objective of the research.

The recommendations of this study with regard to the quantitative aspects are discussed in Chapter VII. This chapter discusses the recommendations to implication which addressed to academia (our fellow researchers), to Companies and business in real estate industry, and to Regulatory authorities and policymakers.

In Chapter VIII, the author provides a summary of the research together with its preliminary thoughts regarding its implications. There is also a discussion on the contributors' specific roles in the overall body of knowledge and the theory. Several potential paths for future research are given below, each of which is led by the findings and context of the present study. Finally, some consideration is given to the shortcomings of this research.

In conclusion, the history of the topic as well as an outline of it have been shown in this study's chapter. The background material identifies the research gap that exists in the literature in a clear and plain manner. The significance of this study can be gleaned from the research problem, as well as the research question, the purpose, and the reason for the investigation. In addition to that, an overview of the investigation, including its conceptual framework, methodological approach, and areas of contribution, is presented in this chapter. Given the structure of this thesis, the next chapter provides a comprehensive discussion of the pertinent theories that surfaced during an in-depth examination of the relevant body of research.

# Chapter II: REVIEW OF RELATED LITERATURE

Real estate investment behavior is a complex phenomenon that has been the subject of numerous studies in the field of finance. However, despite the extensive research on this topic, there is still a need for further investigation into the investment attitudes and behaviors of individuals in different cultural contexts. This study aims to fill this gap in the literature by exploring the proptech adoption in the investment behavior of Vietnamese investors.

This review of related literature will examine previous research on investment behavior with the proptech adoption in Vietnam and other cultural contexts which support the claim that this study has never been done before. The review will highlight the gaps in the literature that the current study aims to address and emphasize the novelty of the current study.

The review will begin by examining previous studies on investment behavior in Vietnam. While there have been several studies on this topic, many of them have focused on specific aspects of investment behavior, such as risk tolerance or investment strategies, rather than providing a comprehensive analysis of investment behavior as a whole. Additionally, many of these studies have been conducted on small samples or specific populations, limiting their generalizability to the broader population of Vietnamese investors.

The review will then examine studies on investment behavior in other cultural contexts. While there is a considerable amount of research on investment behavior in developed countries, there is a lack of research on investment behavior in emerging markets like Vietnam. This is particularly important given the unique cultural and economic factors that may influence investment behavior in these contexts.

Overall, this review of related literature highlights the need for further research on investment behavior in Vietnam and other emerging markets. It emphasizes the novelty of the current study and the ways in which it builds on previous research to provide a comprehensive analysis of investment behavior in the Vietnamese context.

## 2.1. Global real estate market

Real housing prices in the 38 member nations of the Organization for Economic Cooperation and Development (OECD), the club of rich country economies, increased by 16% between the third quarter of 2020 and the third quarter of 2021, data from Refinitiv Oxford Economics show. That's the top rate ever recorded, which dates back 50 years:

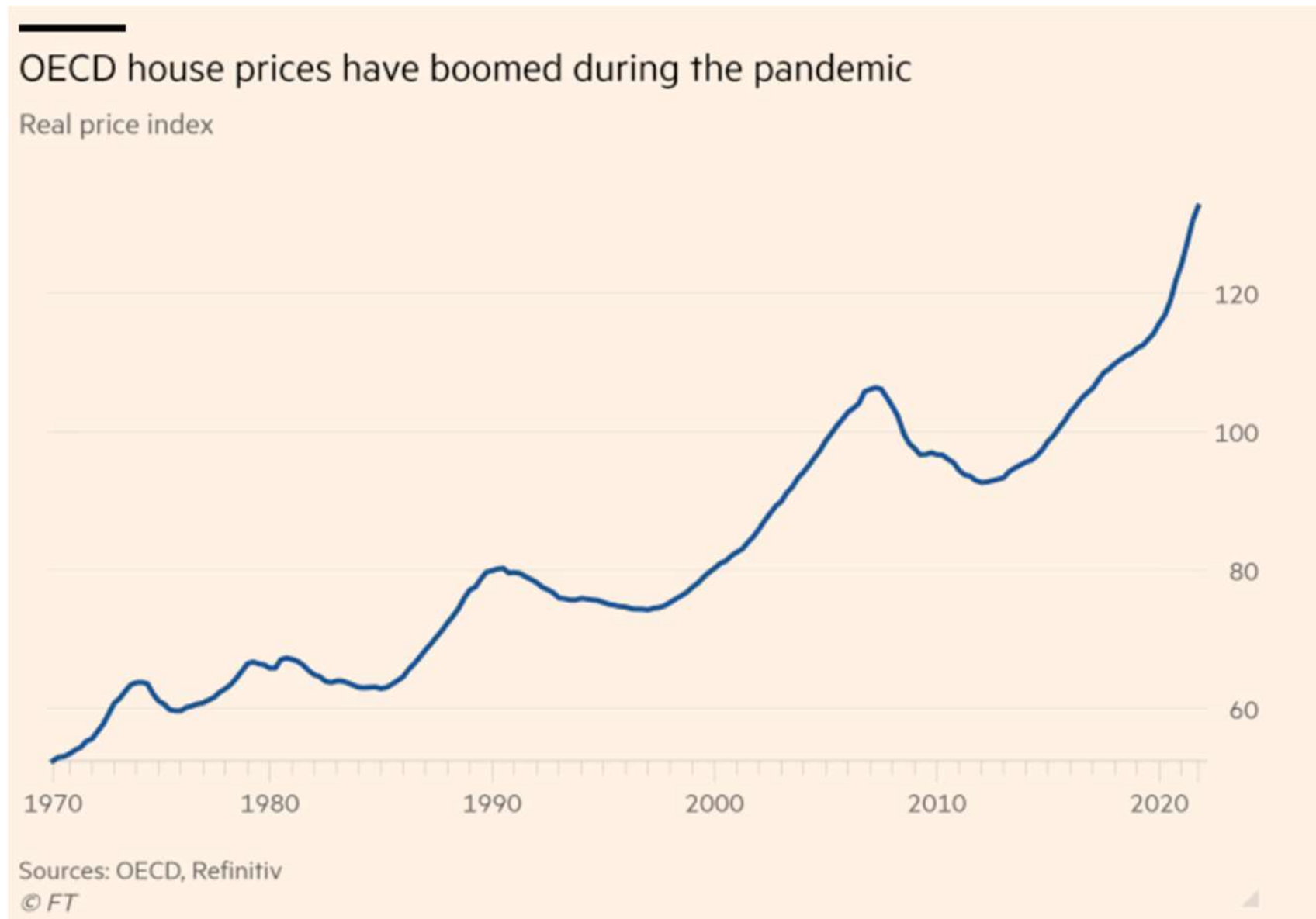

*Figure 2.1.a: OECD house price*

Source: OECD, Refinitiv Oxford Economics

Most analysts believe that this would cause a significant reduction in home price appreciation. As a result of rising mortgage rates and pressure on debt affordability, "we are anticipating house price inflation to slow down in both the US and Europe," says Barbara Rismondo, senior vice president at the rating agency Moody's.

The fast increase in US home prices over the previous two years may "soon flatten out and possibly reverse," according to ING economist James Knightley. Andrew Wishart, the senior property economist at Capital Economics, predicts a 5% price reduction in the UK during the course of 2023 and 2024. If that happened, it would "reverse a fifth of the surge in house prices since the pandemic began," as he puts it.

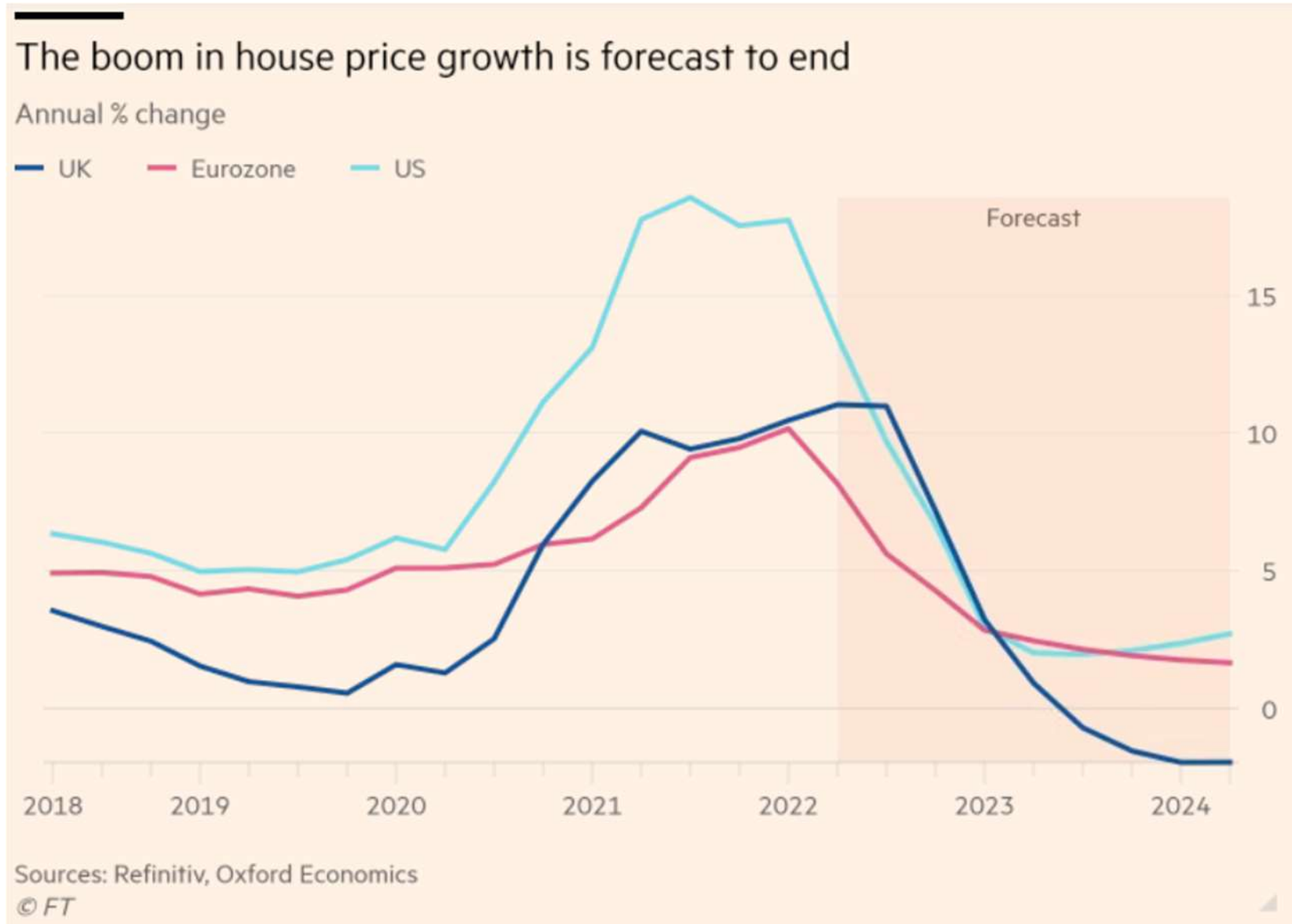

*Figure 2.1.b: The boom in house price growth is forecast to end*

Source: Refinitiv, Oxford Economic

However, few economists anticipate a significant global fall in property prices like that which occurred during the financial crisis of 2008–2009, when economic activity and incomes fell around the world. For five years after the crisis, home prices fell across all OECD countries. It has been seen in OECD countries and particularly in the United States the increase of real estate products that were foreclosed upon and repossessed. Meanwhile, Ian Shepherdson, the chief economist at Pantheon Macroeconomics stated that it's not like what happened in the 2006 crisis.

## 2.2. Today's Property Technologies applied in the property market

PropTech is a far-reaching term used by real estate and technology professionals to cover technologies used for the research, purchase, sale, and management of real property, as well as for companies, particularly start-ups, that specialize in these products and services. (Baum A. , PropTech 3.0: The future of real estate, 2017).

Online home-listing services such as Zillow or Trulia in the United States now allow individual investors and buyers to research all aspects of potential purchases without first engaging a "boots-on-the-ground" agent or broker. Broad access to data previously available only to a select few, coupled with online and proprietary analytical tools, gives investors and buyers the ability to more accurately project various scenarios which leads directly to investing confidence. Likewise, this comparative data arms sellers with tools to track market trends and react to changes in local and broader markets. Another interesting aspect of PropTech penetrates various aspects of property management. For example, Internet of Things (IoT) advances provide managers and owners the ability to remotely video monitor properties. Property managers can use PropTech products to give their owners online portals with real-time access to all aspects of the subject properties, such as rent collections and maintenance status (Mills, 2020). Similarly, PropTech services for investment managers and sponsors to more effectively manage ownership and investment information, effectively allowing individual sponsors to offer services previously only offered by institutional managers. Finally, some jurisdictions are even taking steps to use PropTech for title and ownership records, perhaps the most fundamental aspect of real property owners. One such approach is to use blockchain technologies to track and store property ownership records (Kitajima, 2017).

Although PropTech products and services are flourishing in many other markets, adoption in Vietnam seems to be comparatively slower. Two noteworthy adoptions, however, relate to online-based investing and blockchain-based title records. First, companies are using online platforms to offer individual investors access to institutional-grade real property investments. Previously, purchasing one large building among thousands of individuals would have been impossible for reasons including the inability to offer to and obtain consensus from the investors and the unwieldy management of so many investors. Owners and other real estate stakeholders have tried to implement PropTech to overcome these issues to successfully crowdfund real estate investments in Vietnam. Second, the Vietnamese government is now considering transitioning property records, currently held in paper form and difficult to track and confirm, to a blockchain system. Such a transition would provide a secure and transparent property records system that would lead to investor confidence and likely further stimulate investment.

## 2.3. Real estate stakeholders and their needs

There are multiple stakeholders involved in the real estate industry, however, in line with the definition of smart real estate four stakeholders are of key importance. This includes the end-users or consumers, the agents and their associations (AA), the government and regulators (GR), and the complementary industries (CI) [2]. Table 3 below defines and enlists the basic and other important needs of these stakeholders as identified in the pertinent literature (Ullah F. S., 2019).

| Stakeholders | Definitions | Basic Needs | Other Needs |
|---|---|---|---|
| End-users (Consumers) | This group includes the end-users of the online real estate technologies such as the buyers, renters, or sellers of real estate. These are central to the system and are the main beneficiaries. | Buy/Sell or Rent | 1 Neighbourhood Insights<br>2 Price and Costs<br>3 Mortgages<br>4 Search Tools<br>5 Market Awareness<br>6 Proximity to Amenities<br>7 Layout and Design |
| Government and Regulators (GR) | This includes the government and other regulatory authorities to keep the system transparent and pair. | Regulations and Protection | 8 Economic Growth<br>9 Political Advantages<br>10 Taxes and returns<br>11 Standards Imposition<br>12 Ethical Checks<br>13 Regulations |
| Agents and Associations (AA) | This includes the agents who interact with the consumers for facilitating the buy, sell or rent process. It also includes the associations present for the protection of agents' rights as businessmen. | Business – Profit | 14 Supports to Members<br>15 Networking<br>16 Referrals<br>17 Government Support<br>18 Ethical Checks<br>19 Reputations |
| Complementary Industries (CI) | This includes small and medium industries that are present to help the key stakeholders. | Business – Profit | 20 Networking<br>21 Referrals<br>22 Reputations |

*Table 2: Real estate stakeholders and their needs*

Source: (Ullah F. S., 2019)

According to (Ullah F. S., 2019), it can be seen that the CI and AA have a common goal of excelling in their business and earning more profit. Similarly, the users have the basic need of buying, selling, or renting their properties, and the GR functions to protect the citizens through the imposition of rules and regulations. Further, the end-users or consumers are the people who are at the receiving end of the real estate online technologies and include customers or consumers who utilize these services to buy, sell or rent a property. Their satisfaction is tantamount to the success of the real estate business, therefore, they hold a central position in the hierarchy and all the services revolve around them [7]. Similarly, the AA is the group of business owners and service providers whose aim is to earn profit through business with the end-users or consumers. They play a facilitating role between buyers or renters and sellers and provide their expert services from inspecting the property to listing it online and taking care of the contractual obligations for both parties. In turn, they charge the owners for their services. However, a key concern about the conduct of agents and their abilities to manipulate the situations in favor of the owners, especially in case of a dispute remains a concern for the consumers [8]. This is where the third stakeholder i.e., Government and Regulators step in. This group includes the government bodies and regulatory authorities such as the council and Fairtrade whose main objective is to protect the consumers' rights and make sure that the standards and set of policies related to property deals are followed at all levels. However, the policies are devised in such a way that is optimal for all parties as the agents and owners are also citizens of the state and they have the same rights under the consumer protection laws, so a balance is sought by the GR [9]. In return for these services, the GR taxes all the parties involved in the business and keeps their state of affairs in place. The final stakeholders include the complementary industries that are small and medium-sized industries present to provide support and materials to the real estate businesses, especially the AA. From marketing to specialized web designs and drafting contracts, these CI usually operate on an outsource basis for the AA or owners. Thus, these have similar obligations as the AA and are bound by the GR laws. These CI build upon their networking and reputations to get more business.

## 2.4. Property technologies and their utilization in Real estate

The Big9 technologies include Big Data, AI and Robotics, Cloud, SaaS, IoT, Drones, 3D Scanning, Wearable Technologies, and VR and AR. These have been defined in Table 4 in the following along with their uses in the real estate industry as evident from pertinent literature.

| Big 9 Technologies | Definitions | Uses in real Estate |
|---|---|---|
| Big Data | The Huge or enormous amount of data that cannot be processed or iterated through traditional tools and techniques. | 1. Building Performance Database<br>2. Property Value Analysis<br>3. Crime Rates<br>4. Walkability and Transit Indexes<br>5. Collaborative BEM<br>6. Value Forecasting |
| AI and Robotics | Performing complicated, smart and intelligent functions with computers and minimum human involvement. | 1. Automated Renting<br>2. Fraud Detection<br>3. Business Forecasting<br>4. Blockchain Taxation<br>5. Machine Learning |
| Cloud | Synchronizing data and networking assets over the internet thereby reducing the requirements of placing machinery and computers on site. | 1. Elastic Resource Utilization<br>2. Clients Cross Matching<br>3. Portfolio Management |
| SaaS | Remotely operatable software enabling the users and agents to bring their business to consumers without the need of bringing in computers. | 1. Order Processing<br>2. Resident Management<br>3. Record Keeping<br>4. Building Maintenance |
| IoT | Networked collection of physical devices that can sense the physical aspects of the world such as lightning, exposure to heat and others. | 1. Logistic Synchronization<br>2. Intelligent Community<br>3. Entrance Guard System<br>4. Smart Homes<br>5. Temperature Control<br>6. Home Automation |
| Drones | Unmanned vehicles that can collect accurate and precise data in hard to get around terrains and are operated through remote controls or ground stations. | 1. Aerial Photography<br>2. Volumetric Calculations<br>3. Drone Mappings<br>4. 3D Pictures |
| 3D Scanning | Advanced laser-based scanning devices that enable users to replicate and reproduce models of existing structures in the absence of as-built drawings. | 1. 3D Modelling<br>2. LIDAR<br>3. 3D Images<br>4. Structural Integrity Monitoring<br>5. Point Clouds |
| Wearable Tech | Gadgets based object detection and communication of electronic devices that are integrated into wearable devices or clothing. | 1. Safety Monitoring<br>2. Object Tracking<br>3. Smart Watches<br>4. Body Mediation and Monitoring<br>5. Smart Bracelets |
| VR and AR | Creating virtual worlds or enhancing existing features through digital realities and computer simulations. | 1. 360 Cameras<br>2. VR Headsets<br>3. 4D Advertisements<br>4. 3D Object Recognition |

*Table 2.4.: Real estate technologies*

Source: (Ullah F. S., 2019)

According to (Ullah F. S., 2019), Big data dealing with an enormous amount of data that cannot be handled through traditional approaches has disrupted many fields. In the real estate sector, its application exists in property value forecasting, reality development, city management, and others [10]. AI and Robotics deal with performing complicated and intelligent functions with minimum human involvement and have made their way into the real estate sector through predictive analytics, blockchain taxation, and customer recommendations [11]. Cloud deals with data synchronization and networking over the internet and has applications such as information value for investors' analyses of the real estate market, portfolio management, and elastic utilization of resources [12]. SaaS deals with remote access to software enabling more business and providing more opportunities to real estate agents. IoT deals with networked assets and physical devices and has applications such as Entrance guard systems or smart homes [1]. Drones and unmanned vehicles are displaying huge potential in aerial photography and 3D imagery. Recently wall mounted drones are a new entry into residential real estate markets [13]. 3D scanning that deals with point- cloud-based scanning and data generation has proven its usefulness in real estate through onsite layout plan generation and structural integrity monitoring. Similarly, wearable techs such as smartwatches and smart clothes have huge potential for residents' health and safety monitoring and object tracking during construction stages. Lastly, VR and AR dealing with creating new, or augmenting the existing physical world have huge potential in the real estate sector and can boost the sales process through immersive visualization and interactive playful content [14].

## 2.5. Unified Theory of Acceptance and Use of Technology – UTAUT

Following the study of (Howard, 1969) which provides an explanation for consumer behavior that took into account individuals' rationality, the structure of their decision-making process, and the influence of extraneous factors on their propensity to make a purchase, individuals' reactions to commercial and social stimuli (inputs) promote the choices or purchases (results). It is confirmed that consumers are often motivated to gather information and ask about the goods, perceive risks and costs, and synthesize what they have learned when facing the expectation that is produced by marketing products, such as ease of use, pricing, quality, and performance.

One's inclination to adopt emerges after a thorough examination of all available options has awoken a mental predisposition, which may be positive or negative. That sentiment, together with external (social) and internal (motive, value) factors, is what ultimately decides whether or not someone makes a purchase (Engel, 1986); (Solomon, 2002).

With needs, interests, and values in mind, the level of involvement in decision-making is a reflection of the perceived risk and the priority given to the object of the decision (Assael, 1998; Blackwell, 2008) (Coulter, 2003). Over time, numerous studies have been done in an effort to find the most important elements in the adoption of new technologies.

(Venkatesh V. M., 2003) developed The Unified Theory of Acceptance and Use of Technology (UTAUT) with the intention to incorporate eight prior models of technology acceptance. The resulting research paradigm identified four direct factors which are effort expectancy, performance expectancy, facilitating conditions, and social influence, which determine the usage of a particular technology. (Venkatesh V. M., 2003) demonstrated that the UTAUT was a reliable predictor of technology usage. The results of their study showed that 70% of usage could be determined through the factors identified in the UTAUT. In contrast, the Technology Acceptance Model, one of the eight previous models that influenced the UTAUT, could only predict 40% of the variation in technology usage.

The UTAUT has been utilized since its inception to predict technology acceptance and usage in a variety of fields including education and business. In the business world, studies have focused on many different topics such as mobile marketing (Kiat, 2017), Bitcoin adoption (Silinskyte, 2014), and Internet banking (Rahi S. G., 2018). The version of the UTAUT proposed by (Rahi S. G., 2018) was adopted and modified for use in the current study with permission from the original authors. The quantitative instrument focused on the four direct factors of the UTAUT as well as the intention to adopt internet banking, which was changed to the intention to adopt PropTech in the current study.

## ***Chapter Synthesis***

The literature review has seen the down trend of house price appreciation from the EU market to US and OECD countries, Asian markets also will be affected. Secondly, in developed country like US, we have seen the application of the proptech (technology used in real estate industry) useful and convenient to the real estate users/stakeholders. In comparison, the adoption of Proptech in Vietnam is not as quick as developed countries. Thirdly, this research study reviews some literatures concerning the real-estate stakeholders and the identification of their needs. The proptech technologies and their utilization in Real estate also have been discussed in literature. Finally, the Unified Theory of Acceptance and Use of Technology - UTAUT in short, has been focused for discussion. All the literature summary and future studies of all the literature have been noted in the Appendix.

# Chapter III: THEORETICAL BACKGROUND

## 3.1. Theoretical Framework

The unified theory of acceptance and use of technology (UTAUT) is considered to be the most exclusive theory which incorporates aspects of both the Technology Acceptance Model (TAM), the Technology Readiness Assessment (TRA), and the Motivational Model (Davis, 1992).

Apart from the UTAUT, there are other frameworks, such as:

- Theory of planned behavior (TPB) which is the build-up of different constructs of TAM and TPB (C-TAM-TPB) (Taylor, 1995)
- Model of PC Utilization (MPCU) proposed by (Thompson, 1991),
- Innovation Diffusion Theory (IDT) proposed by (Moore, Development of an instrument to measure the perceptions of adopting an information technology innovation, 1991),
- Social Cognitive Theory (SCT) by (Compeau, 1995).

The UTAUT model suggests the determinants of the behavioral intention to adopt the technology are social influence, facilitating conditions, effort expectancy, and performance expectancy.

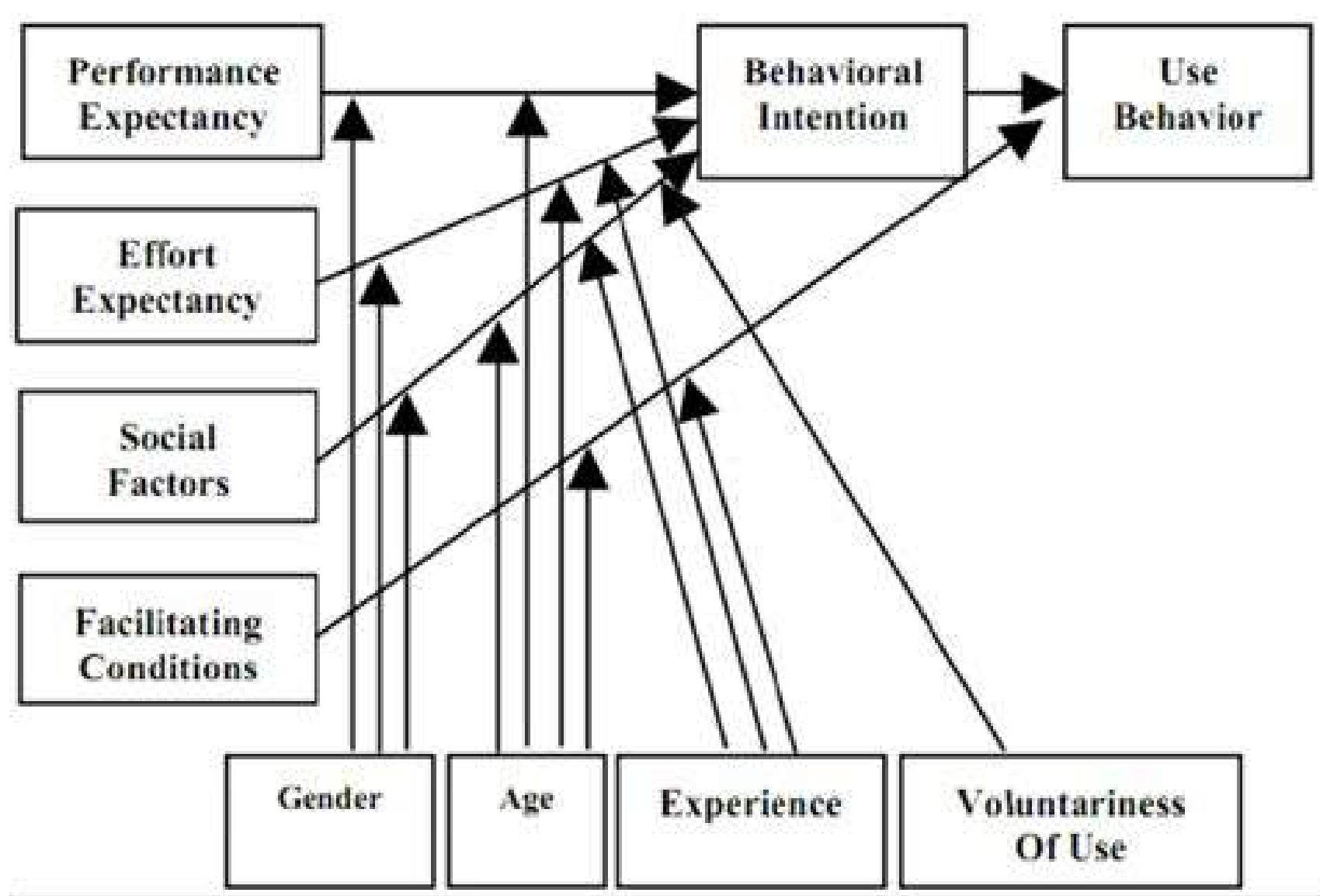

*Figure 3.1.a – **Original UTAUT** suggested by* (Venkatesh V. M., 2003)

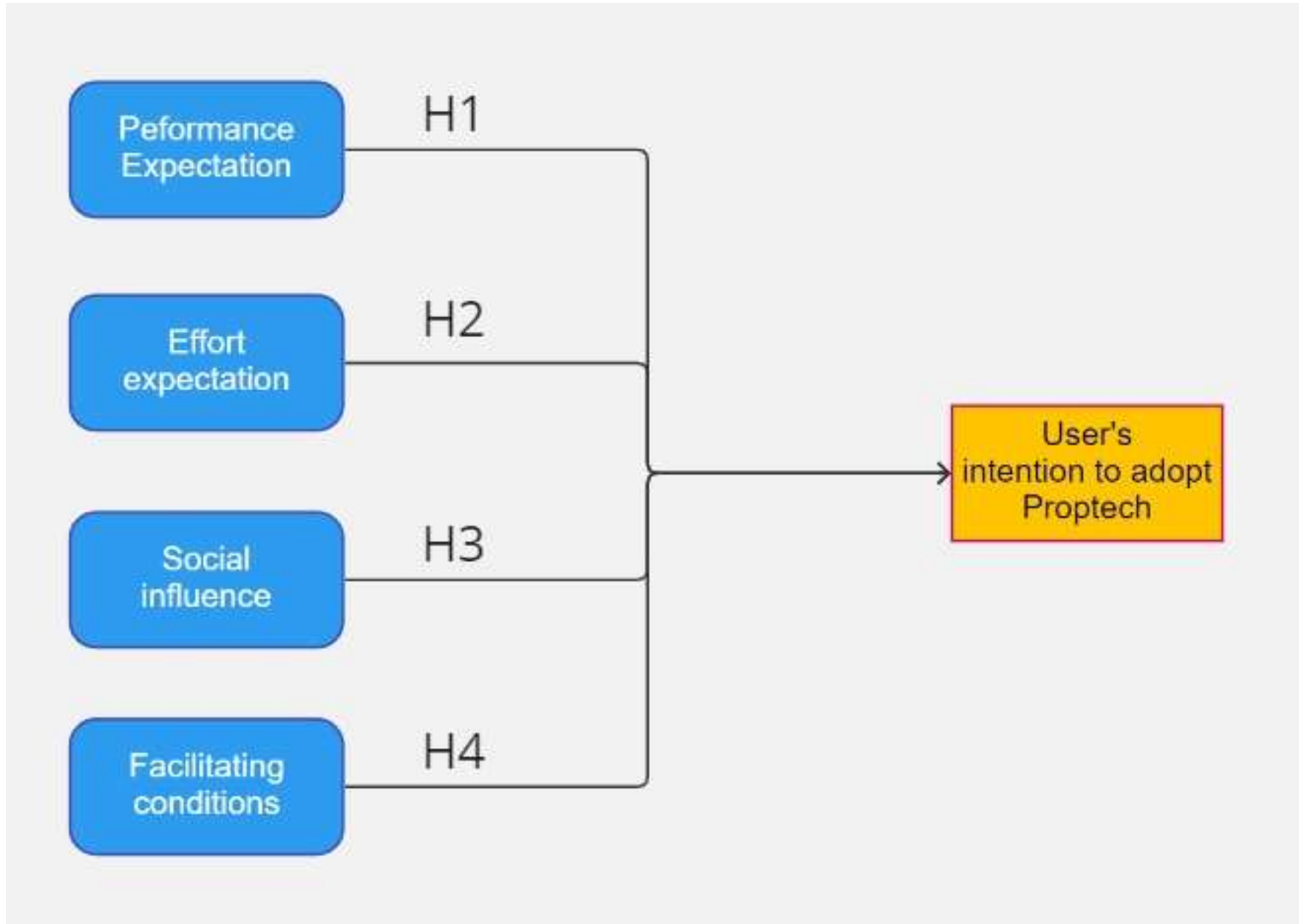

*Figure 3.1.b. – **Proposed conceptual framework***

| | | |
|---|---|---|
| *Performance expectation:* | This measures how much people realize that a system such as Proptech is useful in carrying out their tasks in day-to-day work | (Venkatesh V. M., 2003) |
| *Effort expectation:* | This explains the degree of ease associated with the use of the system. It originates from three constructs of existing models: perceived ease-of-use (TAM/TAM2), complexity (MPCU), and ease of use (IDT) | (Venkatesh V. M., 2003) |
| *Social influence:* | This is defined as the degree of influence that the opinions of others can have on the adoption of a given system. Social influence as a direct determinant of the intention of use is represented as a subjective standard TRA, TAM2, TPB/DTPB, and C-TAM-TPB; social factors in MPCU and image in IDT | (Venkatesh V. M., 2003) |
| *Facilitating conditions* | Facilitating conditions is the perception of support that is provided by institutions and individuals that assist a person in adopting a particular technology | (Venkatesh V. M., 2003) |
| *The behavioral intention of adoption* | This refers to the intention of effective use by the consumer of a future product or service | (Venkatesh V. M., 2003) |

*Table 3.1.* ***- Definitions of the constructs of UTAUT***

Source: (Venkatesh V. M., 2003)

## 3.2. Definition of Terms

It is possible to get further insight from path analysis by merging the ideas presented in Section 3.1 into a competing model (see Figure 3.2.). The alternative model is then proposed once the hypothesized connections have been discussed. The discussion is limited to just four postulated relationships to minimize repetition.

### 3.2.1. Performance Expectancy

(Venkatesh V. M., 2003) describes Performance expectancy (PE) as the degree of performance to which users believe to accomplish their tasks with the help of the system/technology. In this research, performance expectancy is the level to which individuals believe in applying Proptech to achieve his or her goal in real estate operations.

According to (Mills, 2020), thanks to the use of Proptech, users can perceive the improvement in real estate operations. Researchers might compare performance expectancy as extrinsic motivation, relative advantage, and outcome expectancy in other models.

Examples can be found in the sector of internet banking (Alalwan, 2014) postulated that performance expectancy is considered a term of utility encountered during the use of internet banking.

To this end, the thesis research author proposes the following hypothesis regarding performance expectancy:

*H1: Performance expectancy positively influences the user's intention to adopt Proptech.*

### 3.2.2. Effort Expectancy

By definition from (Venkatesh V. M., 2003), Effort expectancy (EE) is the level of ease where user perception of effort excels by the use of Proptech. Effort expectancy in other models was described as perceived ease of use in the Technology Acceptation Model (TAM) and complexity (DOI, MPCU).

According to (Zhou, 2010), there might be a greater opportunity for the user to use Proptech if it is easy to use, and users don't have to pay a lot of time and effort to learn and use.

Significant evidences shown in the studies of (Moore, Development of an instrument to measure the perceptions of adopting an information technology innovation, 1991); (Thompson, 1991) demonstrate the close relationship between effort expectancy on behavioral intention.

When consumers believe that Proptech is simple to use and doesn't need a lot of work on their part, they are more likely to believe that they will achieve the desired results with minimal effort (Zhou, 2010).

Thus, the effort expectancy is hypothesized as follows:

*H2: Effort expectancy positively influences the user's intention to adopt Proptech.*

### 3.2.3. Social Influence

(Venkatesh V. M., 2003) describes Social Influence (SI) as the impact of external variables, such as the thinking from user's social connections like friends or relatives. Social influence in other models is similar to social factors, the subjective norm, and image.

Based on a comparison of different models, we know that the link of social influence will be negligible in a voluntary environment but highly significant in a mandatory one. It has been suggested by (Martins, 2014) that, users' intention to adopt Proptech services is affected by social influence.

Similarly, (Chaouali, 2016) proposed that if a person thinks other influential people approve of the use of a new product or service, he/she is more likely to adopt that product or service.

Therefore, the preceding considerations lead us to hypothesize social influence as follows:

*H3: Social influence positively influences users' intention to adopt Proptech.*

### 3.2.4. Facilitating Condition

As defined by Author (Venkatesh V. M., 2003), "Facilitating conditions (FC)" refer to the effects of the user's knowledge, competence, and resources, as well as the organizational and technical infrastructure, that allows for the successful implementation of Proptech.

Like other constructs of UTAUT, the Facilitation Condition was derived from perceived behavioral control and compatibility in other models. Authors (Venkatesh V. T., 2012) suggested that facilitating condition refers to how consumers feel about the help and support at their disposal to execute a behavioral action.

(Martins, 2014) stated that Proptech's users need a minimum technical ability like computer/mobile phone setup and internet access in order to perform Proptech.

As a result, the relationship of facilitating conditions is proposed as:

*H4: Facilitating conditions positively influence on user's intention to adopt Proptech.*

## ***Chapter Synthesis***

Under investigation of several different technology acceptance models, such as the Technology Acceptance Model (TAM), the Technology Readiness Assessment (TRA), and the Motivational Model (Davis, 1992), Theory of planned behavior (TPB) which is the build-up of different constructs of TAM and TPB (C-TAM-TPB) (Taylor, 1995), Model of PC Utilization (MPCU) proposed by (Thompson, 1991), Innovation Diffusion Theory (IDT) proposed by (Moore, Development of an instrument to measure the perceptions of adopting an information technology innovation, 1991), and Social Cognitive Theory (SCT) by (Compeau, 1995), The unified theory of acceptance and use of technology (UTAUT) by (Venkatesh V. M., 2003) is the most relevant with the determinants of the behavioral intention to adopt the technology are social influence, facilitating conditions, effort expectancy, and performance expectancy. The author of this study proposed conceptual framework based on the UTAUT in Section 3.1. Furthermore, the researcher suggests a competing model and defines each construct of the UTAUT in Section 3.2.

# Chapter IV: METHODS AND PROCEDURES

## 4.1. Research Design

### 4.1.1. Methodological Approach

From the presentation of the proposed competing model earlier, this research recommends an experimental study to evaluate the theoretical relationship from the literature review and it is suggested to test these hypotheses.

According to (Malhotra, 2002), the conceptual framework seeks to quantify the data to explain the causal relationships. Thus, this investigation method is explanatory and driven by the quantitative research tools and techniques.

In order to explore the scientific aspects of the event, conventionally, the quantitative method is considered to apply. This also underlies the deductive model which shows hypothesized relationships.

### 4.1.2. Quantitative Research Design

***a. Sample***

For this preliminary study, a purposive sample was collected from two sources:

- First, a call for participation was placed in a private Facebook group focused on Vietnamese real estate investment, of which the researcher is a member.
- Second, as the researcher is an active real estate investor, he was able to reach out to several contacts in the industry and request participation.

While these methods were sufficient for a preliminary study, a further inquiry would benefit from more robust sampling procedures to ensure the validity and reliability of the data. Eleven investors that fit the criteria of non-Vietnamese individuals who have bought the property for investment purposes in Vietnam or were currently in the research phase of buying property for this purpose filled out the survey instrument.

***b. Participants***

The participants in this study were between the ages of 30 and 56, which suggests that they are likely to be experienced professionals who have been working for a significant amount of time.

This age range also suggests that the participants are likely to have a certain level of financial stability and investment knowledge, which may have influenced their responses to the study.

The fact that the majority of responses came from male investors (70.8%) suggests that men are more likely to be interested in investing than women in Vietnam. This may be due to cultural or societal factors, or it may be related to differences in financial literacy and access to investment opportunities.

All participants identified their nationality as Vietnamese, which suggests that the study was focused on the investment attitudes and behaviors of people within this specific cultural context. This is important because investment decisions are often influenced by cultural and societal factors, and these factors can vary significantly across different regions and countries.

Ten respondents currently live in Ho Chi Minh City (distant/online working), while two lived in Da Nang city. This information provides some context about the geographic distribution of the participants, and suggests that the study may be more representative of the investment attitudes and behaviors of people living in urban areas of Vietnam. It is also worth noting that the fact that some participants were working remotely may have influenced their investment decisions, as they may have different financial priorities and constraints compared to people who work in traditional office settings.

### *c. Quantitative Instrument*

The quantitative survey instrument consisted of 32 items, including an acceptance scale (19 items), usage scale (6 items), and demographics questions (Gender, Age, Location, Education, Occupation, possession of digital devices, and usage of those tools when using PropTech). Likert scales, multi-choice, and open-ended questions were used for the questionnaires.

The acceptance scale was based on the UTAUT model (Venkatesh V. M., 2003) and adapted from a version developed by (Rahi S. G., 2018), which investigated technology adoption in the banking industry.

There are five parts in the questionnaire, including all the constructs below:

i. performance expectancy,

ii. effort expectancy,

iii. social influence,

iv. facilitating conditions, and

v. user’s intention to adopt PropTech.

Permission was gained from the authors to modify their instrument for the current study.

The scale of response was arranged as follows:

(i) Strongly Disagree;

(ii) Disagree;

(iii) Slightly Disagree;

(iv) Slightly Agree;

(v) Agree; and

(vi) Strongly Agree.

Table 4.1.2c shows examples of original and modified items.

The usage items asked participants to indicate the frequency with which they used PropTech to conduct research and gather information, facilitate transactions, and manage their real estate portfolio. The available responses were as follows:

(1) never,

(2) rarely,

(3) sometimes,

(4) often,

(5) very often, or

(6) always.

For every frequency item, there is an open-ended question added to allow participants to develop any ideas on each item.

| Original items | Modified items |
|---|---|
| Internet banking is useful to carry out my tasks | PropTech is useful to carry out my tasks |
| It would be easy for me to become skillful by using Internet banking | It would be easy for me to become skillful by using PropTech |
| People who influence my behavior think that I should use Internet banking | People who influence my behavior think that I should use PropTech |
| A specific person is available for assistance with Internet banking difficulties | A specific person is available for assistance with PropTech difficulties |
| I intend to continue using Internet banking in the future | I intend to continue using PropTech in the future |

*Table 4.1.2c* ***- Original and modified acceptance items***

## 4.2. Data Gathering Procedure

The researcher posted the survey instrument on a public Facebook page dedicated to real estate investing in Vietnam, after gaining the page creator's permission. A few days later, the page creator re-posted the survey to encourage participation. In addition, an email was sent to the researcher's contacts who invest in real estate in Vietnam to ask for their participation. The survey instrument was administered on Google Forms, and the data was downloaded to an Excel spreadsheet. Frequencies and descriptive statistics were calculated for all items.

Data was collected using questionnaires. The survey method used in this research is an online survey to discover user intention to use Proptech. Data collection has been conducted from the end of Jan to early Feb 2022.

The researcher posted the survey instrument on a public Facebook page dedicated to real estate investing in Vietnam, after gaining the page creator's permission. A few days later, the page creator re-posted the survey to encourage participation. In addition, an email was sent to the researcher's contacts who invest in real estate in Vietnam to ask for their participation. Then, the questionnaires under "Facebook's polls" were inboxes to Real estate brokers/consumers and were collected immediately after they were completed.

The author used Google Forms as the survey platform, and the resulting data was then imported into an Excel table. Automatically, descriptive statistics and frequencies were computed for all variables.

## 4.3. Data Analysis Guide

### 4.3.1. Quantitative instrument

The quantitative survey instrument consisted of 32 items, including an acceptance scale (19 items), a usage scale (6 items), and demographics questions (7). The items took the form of Likert scales and multi-choice. The acceptance scale was based on the UTAUT model (Venkatesh V. M., 2003) and adapted from a version developed by (Rahi S. G., 2018), which investigated technology adoption in the banking industry.

The questionnaire items were categorized into the following five constructs: performance expectancy, effort expectancy, social influence, facilitating conditions, and user's intention to adopt PropTech. Permission was gained from the authors to modify their instrument for the current study.

The following options were created for respondents to choose from: (1) strongly disagree, (2) disagree, (3) slightly disagree, (4) slightly agree, (5) agree, or (6) strongly agree.

The usage items asked participants to indicate the frequency with which they used PropTech to conduct research and gather information, facilitate transactions, and manage their real estate portfolio. The available responses were as follows: (1) never, (2) rarely, (3) sometimes, (4) often (5), very often, or (6) always. For each frequency item, an open-ended question was added to give the participants the chance to freely comment and provide more information if they want.

Finally, demographic items enquired about participants' possession of digital devices and usage of those tools when using PropTech in the Vietnamese real estate context. In addition, information regarding age, nationality, gender, country of residence, and Vietnamese language proficiency was gathered.

**4.3.2. Data analysis method and software:**

1. Factor analysis – verify the validity of the variables. (Pett, 2003)
2. Reliability Test – determine the liability of the results obtained from the Likert Scale (Tavakol, 2011)
3. Descriptive Analysis – measure the readiness of Proptech and perceived benefits. (Sekaran U. et al. 2016, p. 43)
4. Regression Analysis – study the significant relationships of the contributing factors. (Mason, 1991)

Previously, the Author has tried to test measurement models and validate the sample for each construct measure. After comparing the different levels of model fit in measurement and SEM models, the Author raises a modified proposed model and it needs to be tested. Respectively, Structural Equation Modeling (SEM) outputs of the modified proposed model are analyzed and discussed in this chapter.

**4.3.3. Hypotheses Testing**

(Kock, 2016) stated that for testing the hypotheses under the circumstance of Partial least squares structural equation modeling (PLS-SEM), we should calculate the P value for each coefficient. Depending on the prior knowledge about the path and sign, the P value may be one or two-tailed.

In order to test all the hypotheses, following the guidance from (F. Hair Jr, 2014), a resample of 5000 has been proceeded.

## ***Chapter Synthesis***

The results from this study have shown a strong support for the unified theory of acceptance and use of technology (UTAUT) findings (Venkatesh V. M., 2003). Since the UTAUT model was created in the western culture, it is valuable and worth trying to test this model in South-East Asian countries. More specifically, this research investigated Proptech by applying the UTAUT model in Hanoi, Vietnam context. It was shown that users' performance expectancy had a substantial relationship with their intentions to use Proptech, meaning that more performance expectancy users have, more likely they are interested in adopting Proptech.

# Chapter V: DATA PRESETATION

## 5.1. Sample Demographics and Data Screening

Demographic items enquired about participants' possession of digital devices and usage of those tools when using PropTech in the Vietnamese real estate context. In addition, information regarding age, gender, location, education, and occupation was gathered.

The author designed a questionnaire, sent it to active customers and staff of the TIEN KHA Real estate brokerage company located in Hanoi, the capital of Vietnam (the biggest real estate market in terms of demographic size), and obtained 142 eligible responses. The author used a structural equation model (SEM) to examine the data and test the hypotheses which combine all latent variables. Real estate agents/brokers and potential customers are the targeted populations of interest. Real estate agents/brokers working in both the private and government sector in Hanoi, mainly in the center of Hanoi (inside of ring road no 3) constitute the sampling frame for this research. Because of the frequent use of the internet, social network, and mobile phone, Real estate agents/brokers are chosen as appropriate as an adequate sampling frame for the research.

## 5.2. Measurement Development

For the purpose of the research survey, a questionnaire was designed by drawing constructs and items from the relevant literature. The questionnaire was made and carried out in the native Vietnamese language throughout its entirety. The questionnaire for the survey is broken up into two sections:

- The questionnaire's first section includes demographic profiles of the participants.
- While the second section of the questionnaire contains measurement items that were adapted from (Venkatesh, 2012) and users' intentions to adopt Proptech. These items measure users' performance expectancy, effort expectancy, social influence, facilitating condition, and overall intention to adopt Proptech (Rahi, 2017). Each measurement item is based on a Likert scale: (i) Strongly Disagree; (ii) Disagree; (iii) Strongly Disagree; (iv) Slightly Agree; (v) Agree; and (vi) Strongly Agree (Joshi, 2015). Then, the Author evaluates the measurement model for construct reliability, indicator reliability, convergent validity, and discriminant validity.

To assure reliability, Cronbach's (α) is recommended and Composite Reliability (CR) is also preferred (Henseler, 2009).

## 5.3. Structural Model Testing

Following the instruction from (Hoyle, 1995), the author of this research use Structural equation modeling (SEM) to test the model and assess the complex relationship between all the latent variables including performance expectation, effort expectation, social influence, facilitating condition, and the dependent variable behavioral intention to adopt Proptech.

The examination of indicator loadings is to confirm the convergent validity. Fig. 9 displayed results where, a threshold level of 0.6, as recommended by (Chin, 1998), represents the support of factor loading values. Because all the values were above 0.6, convergent validity is verified.

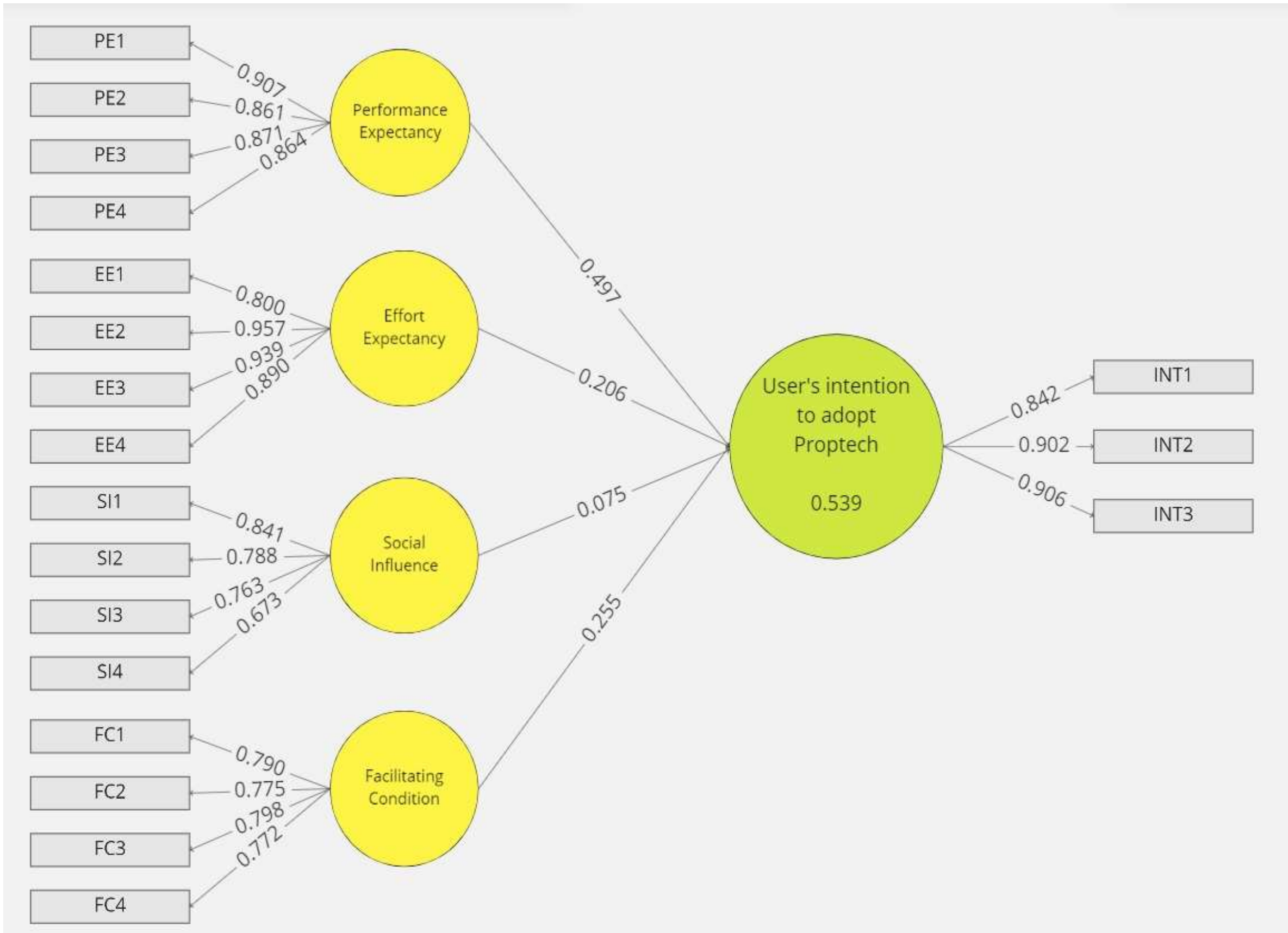

*Figure 5.3. – **Measurement Model***

In addition, the convergent validity of the measurement model is also evaluated by calculating the average variance extracted and the composite reliability.

The average variance extracted is represented by the total amount of variance in the indicators which accounted for latent construct.

Furthermore, the composite reliability of this measurement model should also be evaluated. Hence, the estimation of the average variance extracted (AVE), proposed by (Fornell, 1981), by recommended values because it should be more than 0.5 in value.

(Hair J. F., 2011) suggested that values for the composite reliability (CR) greater than 0.7 where the construct indicator supports the validity of the latent construct.

The table on the next page narrates the results of the measurement model:

*Table 5.3.* ***– Results of Measurement Model***

| Constructs | Loading | (α) | CR | AVE |
|---|---|---|---|---|
| **Performance Expectancy** | PE | 0.899 | 0.930 | 0.767 |
| Proptech is useful to carry out my tasks. | 0.907 | | | |
| I think that using Proptech would enable me to conduct tasks more quickly. | 0.861 | | | |
| I think that using Proptech would increase my productivity. | 0.871 | | | |
| I think that using Proptech would improve my performance. | 0.864 | | | |
| **Effort Expectancy** | EE | 0.919 | 0.944 | 0.808 |
| My interaction with Proptech would be clear and understandable. | 0.8 | | | |
| It would be easy for me to become skillful by using Proptech. | 0.957 | | | |
| I would find Proptech easy to use. | 0.939 | | | |
| I think that learning to operate Proptech would be easy for me. | 0.89 | | | |
| **Social Influence** | SI | 0.774 | 0.852 | 0.591 |
| People who influence my behavior think that I should use Proptech. | 0.841 | | | |
| People who are important to me think that I should use Proptech. | 0.788 | | | |
| People in my environment who use Proptech services have a high profile. | 0.763 | | | |
| Having Proptech services is a status of symbol in my environment. | 0.673 | | | |
| **Facilitating Condition** | FC | 0.791 | 0.864 | 0.615 |
| I have the resources necessary to use the Proptech. | 0.79 | | | |
| I have the knowledge necessary to use the Proptech. | 0.775 | | | |
| Proptech is compatible with other technologies I use. | 0.798 | | | |
| A specific person is available for assistance of Proptech difficulties. | 0.772 | | | |
| **User's intention to adopt Proptech** | INT | 0.859 | 0.914 | 0.781 |
| I intend to continue using Proptech in the future. | 0.842 | | | |
| I will always try to use Proptech in my daily life. | 0.902 | | | |
| I plan to continue using Proptech frequently. | 0.906 | | | |

(Hair J. R., 2009) stated that the combination of measurement and the structural model helps to form the structural equation model (SEM). The proportion of variability (R2) of the dependent construct behavioral intention was assessed after the examination of the relationships reflected in the theoretical model.

Partial least squares (PLS) were used as the approach to estimate the SEM since it is more adaptable following the study of (Chin, 1998). In order to verify the validity of the proposed theoretical model, the author uses the Smart PLS 3.0 software.

The researcher followed the recommendation from (Henseler, 2009) which divides the process into two phases:

Firstly, the composite reliability should have values greater than 0.7 to verify the reliability of the internal consistency of the constructs.

Secondly, convergent validity is then applied to examine the unidimensionality of the construct by means of average variance extracted (AVE) which should have a value greater than 0.5.

## ***Chapter Synthesis***

In this chapter, the author presents the sample demographics and data screening methods. Real estate agents/brokers are chosen as appropriate as an adequate sampling frame for the research. For the purpose of the research survey, a questionnaire was designed by drawing constructs and items from the relevant literature. The Structural equation modeling (SEM) has been used to test the model and assess the complex relationship between all the latent variables including performance expectation, effort expectation, social influence, facilitating condition, and the dependent variable behavioral intention to adopt Proptech.

# Chapter VI: FINDINGS AND IMPLICATIONS

## 6.1. Specific Findings

Participants ranged in age from 30 to 56. The majority of responses came from male investors (70.8%). All participants identified their nationality as Vietnamese. Ten respondents currently live in Ho Chi Minh City (distant/online working), while two lived in Da Nang city.

Table 6.1. illustrates the sample demographics. Most of the participants were male (71.13%) while females were (28.87%). The age of the participants 0% is for less than 20 years, 20.42% counts at the age between 21 to 30 years, 63.38% for 31 to 50 years, 22.87% for those participants who have age between 51 to 60 years, 2.82% was participants having age above 60.

The study also assessed the education level of the participants. Results showed that most of the participants had graduate level qualifications (n=70, 49.3%) followed by those who had attended college qualifications (n=60, 42.25%). The number of respondents who had attended high school was only 3 while the participant who qualified for high school were none. These findings revealed that the educated respondents were more inclined toward Proptech as compared with uneducated respondents (below High School, 2.1%).

This study has revealed the occupation of the respondents. The results showed that the highest number of respondents were employed (n= 106, 74.65%) followed by Self-employed/freelancer participants (n=31, 21.83%). There were only 3 respondents with pensioner titles. The respondents with a student profile had the second lowest number (n=2, 1.41%) while no respondent is unemployed.

| Demographic | Category | Frequency (n=142) | Percentage (%) |
|---|---|---|---|
| Gender | Male<br>Female | 101<br>41 | 71.13%<br>28,87% |
| Age | < 20<br>21-30 | 0<br>29 | 0<br>20.42% |
| | 31-50<br>51-60<br>> 60 | 90<br>19<br>4 | 63.38%<br>22.87%<br>2.82% |
| Education (Level) | Below High School<br>Attended High School<br>Attended College<br>Graduate<br>Post Graduate | 0<br>3<br>60<br>70<br>9 | 0%<br>2.1%<br>42.25%<br>49.3%<br>6.34% |
| Occupation | Student<br>Employed<br>Self-employed<br>Unemployed<br>Pensioner | 2<br>106<br>31<br>0<br>3 | 1.41%<br>74.65%<br>21.83%<br>0%<br>2.11% |

*Table 3.1.* ***- Demographic Profile of the Respondents***

## 6.2. Hypothesis Results

The table below shows the Partial least squares (PLS) estimation results:

| # | Construct | β | S.E | t-values | P-Values | Results |
|---|---|---|---|---|---|---|
| H1 | PE -> INT | 0.497 | 0.048 | 10.316 | *** | Supported |
| H2 | EE -> INT | 0.206 | 0.047 | 4.384 | *** | Supported |
| H3 | SI -> INT | 0.075 | 0.032 | 2.324 | ** | Supported |
| H4 | FC -> INT | 0.255 | 0.043 | 5.950 | *** | Supported |

Note: Significance level where, *p < 0.05, **p < 0.01, ***p < 0.001.

*Table 6.2.* ***Hypothesis Testing***

**Source:** (F. Hair Jr, 2014)

The findings of the structural model demonstrate strong correlations relationships between each of the four hypotheses and the respectively endogenous variables, explained as follows:

H1: (β = 0.497, p< 0.000) indicates that there is a strong connection between performance expectancy and users' intention to adopt Proptech.

H2: (β = 0.206, p< 0.000) represents the positive relation of effort expectancy with the user's intention to use Proptech.

H3: (β = 0.075, p< 0.001) supports the social influence's positive relationship with the user's intention to use Proptech.

H4: (β = 0.255, p< 0.000) showed that facilitating conditions have a strong influence on the user's intention to adopt Proptech.

## 6.3. Research Objectives (Answers)

### 6.3.1. Research Question 1 (Acceptance)

***Performance Expectancy (PE)***

Performance expectancy is the degree to which a person supposes that a particular technology will assist in the achievement of a particular goal (Venkatesh V. M., 2003). This subscale displayed a high mean value (M = 4.42). A majority of participants (53.5%) agreed or strongly agreed that PropTech would be useful in carrying out investment tasks and would increase their productivity. A big majority (73.9%) imagined that PropTech would help them complete tasks at a quicker pace. These responses are most likely indications of positive experiences using technology for business tasks, including real estate investing. See the Figure below:

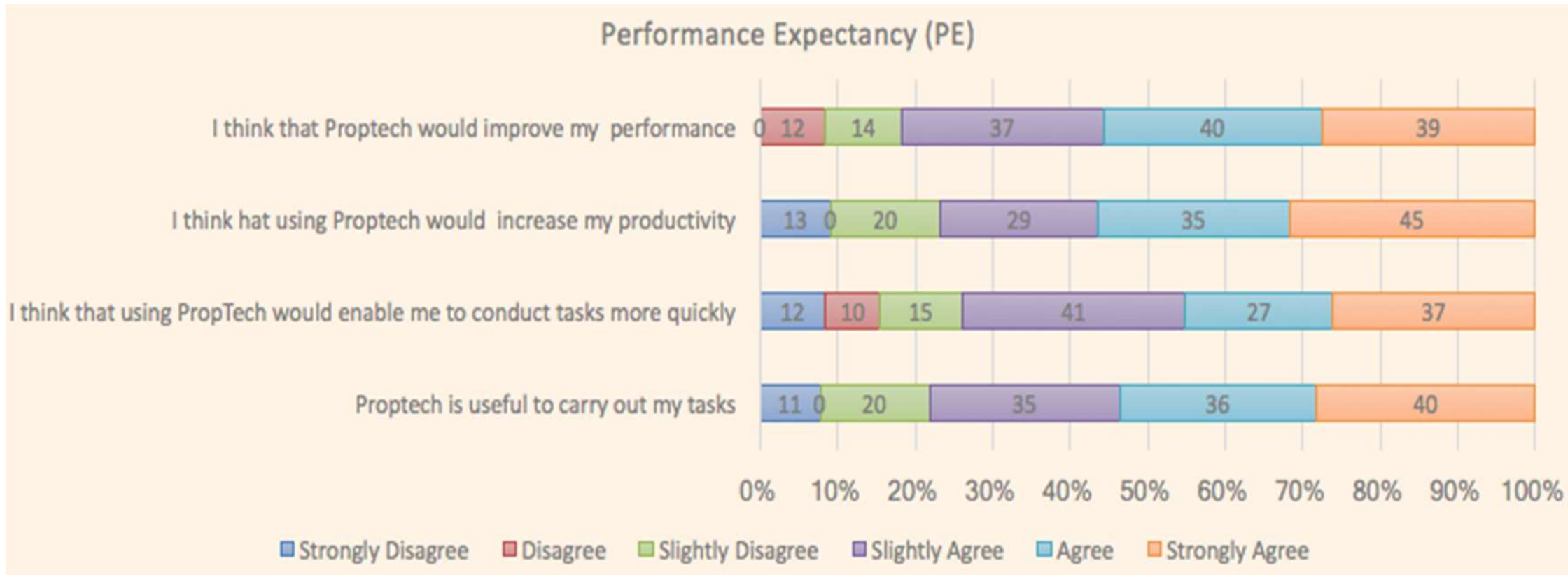

Figure 6.3.1a – Performance Expectancy (PE)

Source: Author

***Effort Expectancy (EE)***

Effort expectancy is how easy a person believes it will be to use a particular technology (Venkatesh V. M., 2003). The highest mean value was associated with this subscale (M = 4.56). A bigger proportion of respondents agreed or strongly agreed that PropTech would be easy to use (62.7%) and over 80% believed that learning PropTech would be easy for them. However, less than half of the participants (45.07%) agreed or strongly agreed with the statement, "My interaction with PropTech would be clear and understandable." These responses are not surprising because all the participants who were surveyed are technologically proficient professionals who seem to make extensive use of technology in their work and investments. See the Figure below:

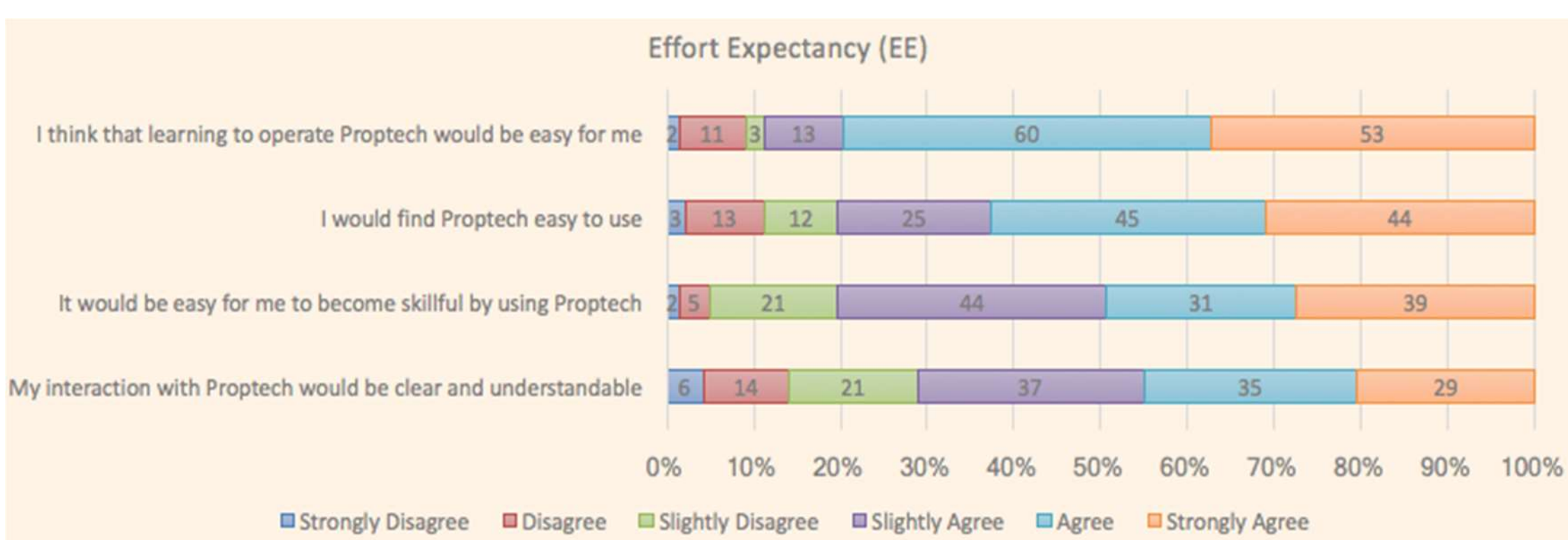

Figure 6.3.1b – Effort Expectancy (EE)

Source: Author

***Social Influence (SI)***

(Venkatesh V. M., 2003) defined that social influence is the level at which individual(s) believes others (family, relatives, friends) want them to use the system or technology.

The lowest mean value was found in the social influence subscale (M = 3.99). Furthermore, 45.77% of participants agreed or strongly agreed that the usage of PropTech was a status symbol among their peers. In addition, 34.51% of participants believed that people who influenced their behavior wanted them to use PropTech. This may be explained by the current lack of social communities for real estate investors in Vietnam, especially outside of large cities. Online investor communities, such as Propzy, have provided some support for investors but are still not reachable to Vietnamese investors. Without a network of peers, real estate investing remains a largely solo activity in Vietnam, especially for investors. See the Figure below:

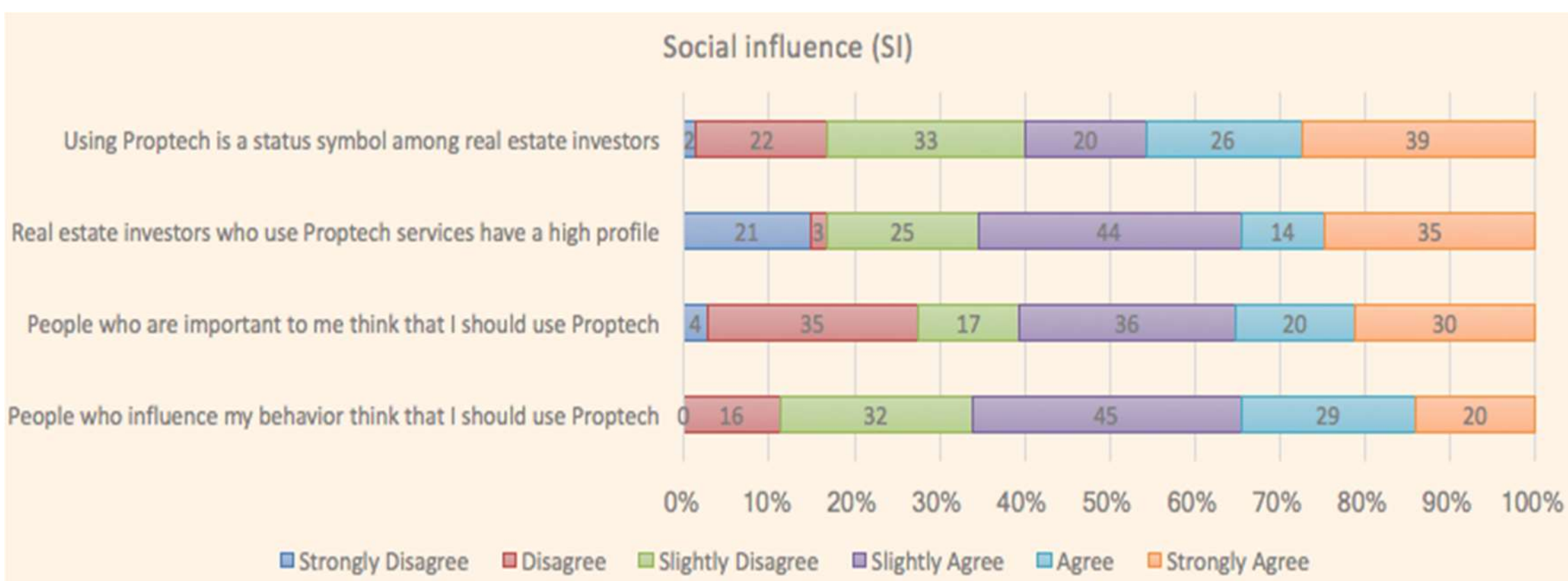

Figure 6.3.1c – Social Influence (SI)

Source: Author

***Facilitating Conditions (FC)***

Facilitating conditions are the perception of support that is provided by institutions and individuals that assist a person in adopting a particular technology (Venkatesh V. M., 2003). A majority of participants (59.86%) agreed or strongly agreed with the statement, "I have the resources necessary to use PropTech." This is most likely true because the participants use these technologies effectively in their daily activities and are only not using Vietnamese PropTech because of the lack of available products. See the Figure below:

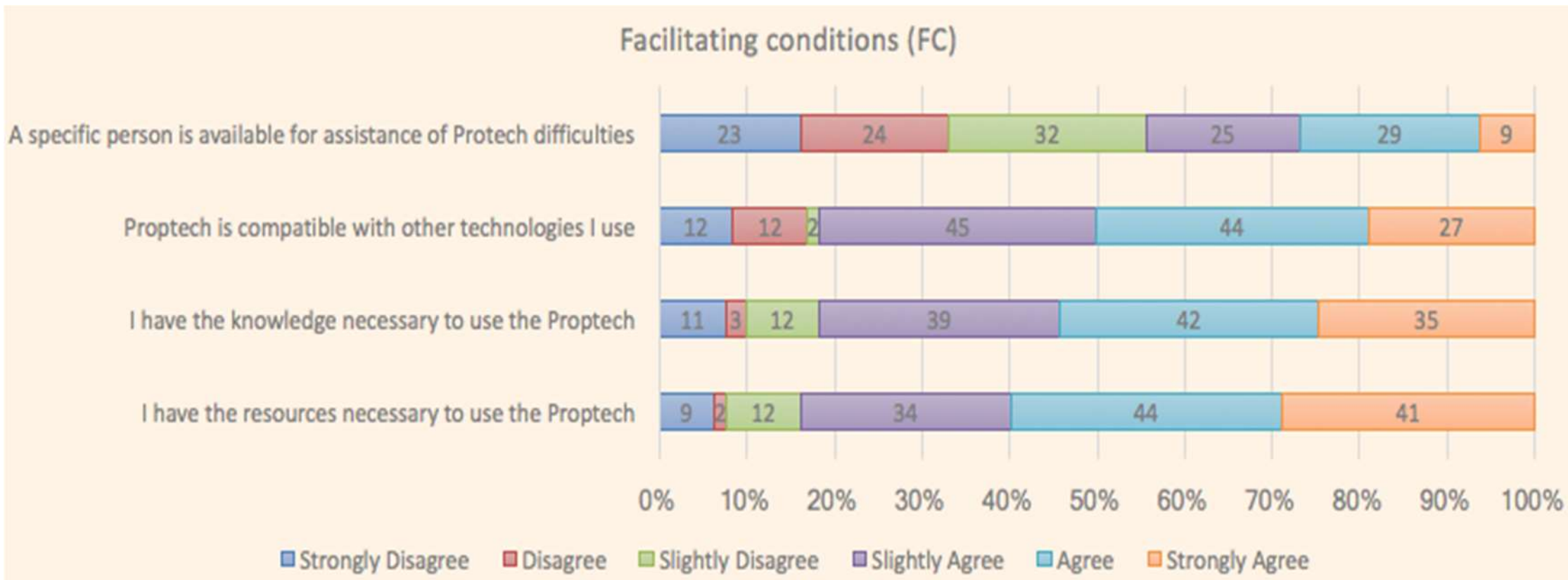

Figure 6.3.1d – Facilitating Conditions (FC)

Source: Author

**<u>User's intention (UI)</u>**

The responses in the subscale were positive (M = 4.41). More than half of the respondents agreed or strongly agreed with all questions. The highest rated items were associated to use PropTech in the future and using it frequently. This demonstrates a real desire on the part of the participants to continue using the tools they are already employing and the possibility that they would be eager to make use of future technological developments in the market. See the Figure below:

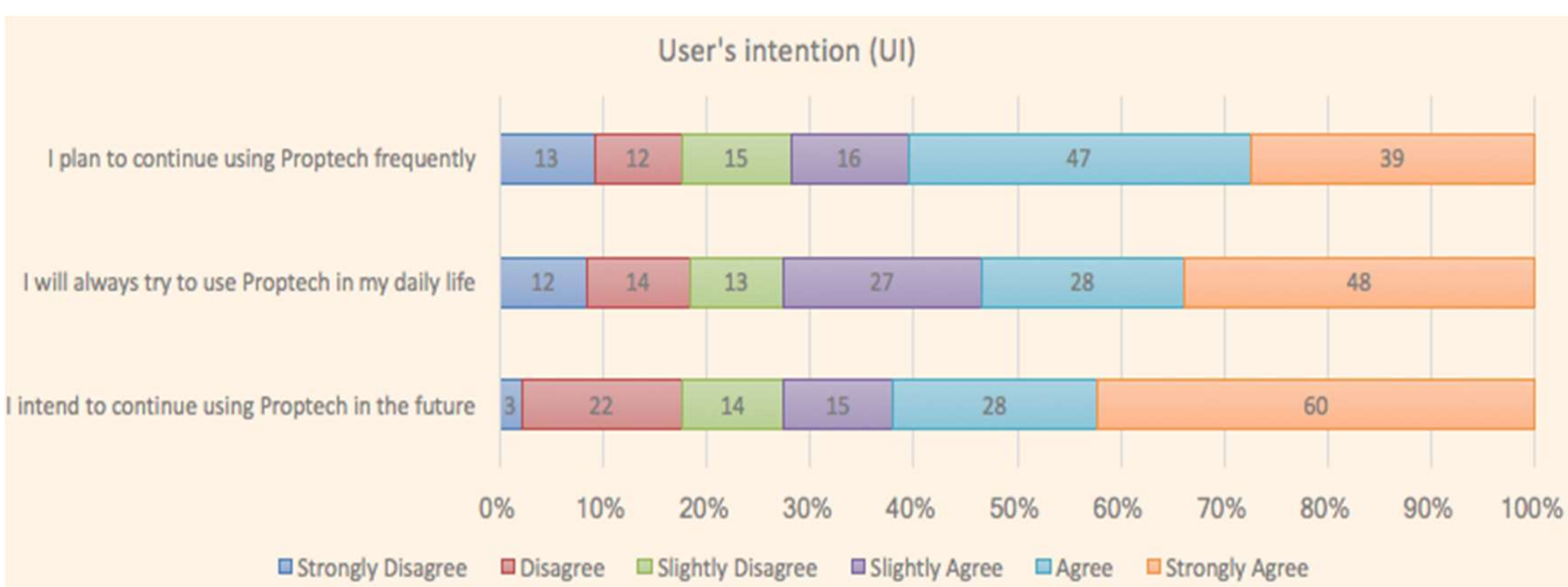

*Figure 6.3.1e – **User's intention (UI)***

Source: Author

#### 6.3.2. Research Question 2 (Usage)

The data regarding the usage of PropTech showed that participants mainly utilized the technology for gathering information on properties and markets (M = 4.14) rather than transactions (M = 3.83) or portfolio management (M = 3.98). This makes sense because investors, in all markets, spend the majority of their time researching markets and analyzing deals often analyzing 100s of potential deals before taking action. In addition, if a real estate agent and property manager are employed by the investor, most of the tasks related to the transaction and portfolio management will be taken care of by those professionals.

When queried regarding the specific usage of PropTech for doing research and gathering information, respondents stated that they utilized technology for a variety of tasks. Several participants mentioned Google as a valuable tool for communication (Gmail) and also to keep records and create documents (Google Drive). One feature of creating documents such as spreadsheet analyses of properties in Google Drive was the ease with which these documents could be shared with others. One participant also mentioned using Google's search function to find contact information for companies and individual property owners. Furthermore, Google Maps was identified as a useful tool to look at the subject property and surrounding area.

Many participants mentioned that blogs, online news sites, and virtual communities were important resources for investigating specific markets. According to the sharing of one participant when talking about a particular online investment community, "This community is full of investors willing to share and learn, and it's where I connected with my REI (Real Estate Investing) partners. This online community spring boarded my investments on both sides of the Pacific." Finally, Vietnam-specific property portals such as Propzy and Batdongsan.com.vn were used by participants to find properties and search for market comparable. Despite this, there is a possibility for development with each of these gateways. One respondent stated that she/he was "finding a weakness in PropTech for well-run real estate portals at this time."

Concerning transactions, cloud-based services, like Google Drive, were mentioned again for their ability to "receive, analyze, and confirm data, review documents, store and share them." In addition, payment services like Vnpay for credit card payments and Momo were mentioned. For management, cloud-based services were the most used technology followed by email, which was used to receive monthly statements and communicate with property managers.

### 6.4. General Findings

Acquisition and management of the portfolio in Vietnam is very analog and utilizes very little if any PropTech. Acquisitions were all facilitated with traditional connections formed face-to-face. Management uses local companies that provide written reports but no owner portal or similar service.

Both interviewees highlighted the lack of availability of PropTech resources in the Vietnamese market. While their opinions cannot be seen as representative of the sample, their experience and knowledge do show that this is an issue worth pursuing in further research.

Furthermore, some interviewees stated that it was hard to comment due to their lack of usage of dedicated PropTech systems. However, the dearth of technology usage in this space does present an opportunity for Vietnamese PropTech firms to develop products or at least make current products accessible. Developments in PropTech like property portals, social media sites, and management software that give Vietnamese property managers a way to share information would all be welcome additions to the market.

## ***Chapter Synthesis***

The study assessed sample demographics, age and also the education and occupation of the respondents. The study showed that educated respondents were more inclined toward Proptech as compared with uneducated respondents. When testing the hypotheses, the findings of the structural model demonstrate strong correlations relationships between each of the four hypotheses and the respectively endogenous variables.

Most of the questionnaire participants consider Proptech as an useful service. They thinks that Proptech is easy to use and the usage of PropTech was a status symbol among their peers. They have all the conditions to use Proptech and will more apply Proptech in every day work.

For Proptech usage, that participants mainly utilized the technology for gathering information on properties and markets rather than transactions or portfolio management. Google maps, Gmail and Google drive are very helpful in the work. Related to real estate transaction, very little of participants mentions to use Vnpay and Momo (Electric wallet) for the transaction.

In General, Vietnamese real estate stakeholders are slow to adapt Proptech into the everyday uses. As an important aspect of the Vietnamese character, people are very traditional and that is why they'd prefer to meet up face-to-face to share and discuss information.

# Chapter VII: RECOMMENDATIONS

## 7.1. Addressed to Academia

This research provides a contribution to the academic community in the sense that it incorporates several most widely used constructs in recent literature in the same model to predict the adoption of Proptech services, from the point of view of a group of Vietnam Proptech stakeholders. Consequently, a new model can be utilized in different contexts.

## 7.2. Addressed to Companies, Businesses in real estate industry

Relating to the practical contributions, this thesis report provides useful information for the business sectors to examine the market reaction before possible implementation of the newly related service. Furthermore, it helps business sectors to promote segmentation and communication strategies based on the discovery of the factors that precede the intention to adopt Proptech.

According to another study of the data, there are two separate types of the user profile. This leads us to the conclusion that it is essential to design solutions that are comprehensive and risk-free for both types of users, most likely making use of specialized technological resources.

Users of proptech have access to a variety of benefits, including the possibility of saving time, increased convenience, and the introduction of novel consumer experiences. On the flip side, everyone who is involved in the process of the transaction, including the end-user (consumers), the government and regulators (GR), agents and associations (AA), and complementary industries (CI), can benefit from the provision of this new service (Ullah F. S., 2019).

## 7.3. Addressed to Regulatory authorities and policymakers

Because of regulatory factors, the real estate transaction process in Vietnam is still considered to be very complex. That is why we should highlight this complexity with technological solutions.

The growth of proptech in Vietnam is contingent on a knowledge of the features of this market as well as the preparation of the internal capacity of businesses that are interested in engaging in this line of work. Awakened by the high penetration rate of proptech, real estate corporations have expressed interest in assuming a major role in this paradigm.

## ***Chapter Synthesis***

This section enlists the recommendations of the study. The purpose is to offer ideas on how the findings of this study can be implemented in academia to further this field of study. It also offers suggestions on how the challenges of the industry or individuals can be addressed for better outcomes. Last but not least, the findings of this research which have been mentioned in Chapter VI and previously summarized also have some theoretical and managerial/practical implications to the Government (regulatory authorities and policymakers) to suggest a framework to improve and/or a roadmap/strategic plan for long-term action plan.

# Chapter VIII: SUMMARY AND CONCLUSION

During the course of developing the research thesis, the Author has conducted an analysis of the intention to adopt Proptech in a developing country context. This brings together, in a fresh research context, the theoretical rationale underlying a number of fundamental economic theories. The central research question underpinning this thesis was:

"How do users in the Vietnamese real estate market perceive and utilize Proptech to research, acquire, and manage their holdings?"

The primary objective of this study was to develop and validate a conceptual research framework that incorporates the adoption of Proptech based on relevant theories and the literature.

## 8.1. Summary of the Report

All pertinent guidelines towards identifying the literature review and a full evaluation of potential theories and theoretical literature were undertaken in order to answer the research question and accomplish the study objective in Chapter II. Chapter III developed a quantitative research model based on merging the review stream. Due to the fact that this study aimed to analyze causal events, a competing quantitative research approach was incorporated into Chapter IV. In Chapter V, the quantitative models were evaluated and compared with the main data that was collected, and the results of those tests and comparisons are discussed. In addition, the investigation's findings as well as a discussion may be found in Chapter VI. The recommendation to implication is shown in Chapter VII. This chapter provides a summary of the findings in an effort to address the theoretical and practical consequences as well as the contributions made by the study. The chapter comes to a close with a discussion of the constraints imposed by the study as well as future studies' direction.

## 8.2. Overall contributions

The major areas of the research's overall contribution are:

The conceptual architecture proposed is confirmed at a high level of explanatory and achieved a good fit to the data. As a consequence, this research adds to the existing body of knowledge by providing empirical support for a large number of explanatory variables in the comprehensive model.

Related to managerial implications, the research makes a substantial contribution to the existing body of knowledge by combining information data from a developing country into empirical generalizations of the findings.

Last but not least, despite the fact that this research have made modification on the construct measures and new measures have been introduced, in order to confirm their reliability and validity, a proper methodological process has been implemented following the guidance from (Churchill, 1991). Hence, this study has brought contribution to international business literature a quantitative methodological approach.

## 8.3. Limitation

Similar to other studies, this research has limitations which are discussed as follows.

Firstly, although a non-linear scale for questionnaire responses has been utilized, the dynamic connection between the intra-relationship and the barriers was not analyzed. In order to improve the overall quality of the study, this could be investigated in a future study in which we can confirm the dynamics of the system.

Secondly, Vietnam is a member of the Association of Southeast Asian Nations (ASEAN) countries which consists of many other developing countries. This current research only focuses on analyzing the context of Vietnam real estate sector. The comparison with other countries in the region or other continents would be very helpful to the literature if we take consideration on rapid changing in smart real estate technology in the era of industry 4.0.

Thirdly, we can see a limitation on the sample of Proptech users which was, conveniently, restricted to real estate/brokers and customers of a single real estate company in Hanoi City, one of the largest markets in Vietnam. This might pose some bias of viewpoints that were systematic in nature. In the future, it is recommended to look further into other results from across all the provinces of Vietnam in order to get better insight and comparison.

In addition to this, there is a possibility of utilizing additional pertinent aspects that improve the adoption model that is being used.

However, despite the fact that there were so many limitations, the findings of the research are in line with what the author had anticipated as well as the studies that were used as references.

## 8.4. Future direction

While focusing on the analysis of the impact of the factors suggested by (Venkatesh V. M., 2003): performance expectation, effort expectation, social influence, and facilitating conditions, this research did not investigate how the value is created with the help of Proptech. There are a lot of benefits from the value creation to house owners/tenants in the Vietnamese real estate sector. In the future, researchers can also study other aspects such as IT security, economic aspects, law, and regulations, or personal integrity with the adoption of Proptech.

In addition, the digital infrastructure (software or hardware designs) for Proptech implementation is another potential topic for future research. In this regard, it would be instructive to investigate whether digital solutions and services are technically and economically feasible to implement. The innovative effort related to digitalization in Vietnam's real estate sector will also benefit from exploring new and existing business models.

The results of this research can be double-checked with samples from other regions or real estate companies that have seen or not Proptech deployment. Comparative studies can use a complete model with the perceived cost construct to examine the outcomes associated with this variable.

Future studies also can use the model proposed on other segments or another group of real estate stakeholders like non-Vietnamese investors' intention to use Proptech for comparison.

The current study did offer some insight into possible avenues of PropTech development such as property portals, social media sites, and bilingual property management software; however, further research must be conducted to discover the specific needs of the end users. This is because the study did offer some insight into possible avenues of PropTech development. This type of research would address the limitations of the preliminary study that was given here by collecting data from a greater number of participants and adopting sampling techniques that were more reliable. Researchers and developers of PropTech can answer the expectations of the growing number of foreign investors who regard Vietnam as an important property market by expanding on the data presented in this chapter.

In order to learn more about the elements that are driving consumers' intention to use Proptech, the future study can consider to employ other potential variables like e-customer service, assurance and reliability etc. to expand the UTAUT model.

## ***Chapter Synthesis***

The purpose of this preliminary study was an attempt to examine real estate stakeholders' acceptance and usage of PropTech in the Vietnamese real estate market with a small group of participants. The results showed that the participants were generally accepting of PropTech in this context and used some aspects of it in their daily investing activities. PropTech was mainly employed to research potential markets and properties. Websites such as batdongsan.com.vn and nhatdatonline.vn, which allow users to search for properties, were accessed by the participants, but these sites are targeted at local investors and often do not provide bilingual services. It may not come as a surprise then that among the sample the most used technologies were supplementary applications like the Google Suite rather than any products developed specifically for real estate investing in Vietnam. In addition, the social media site Facebook was identified as a valuable resource to meet and communicate with other investors, as well as to research and gather information. However, this social media tool is focused mainly on the US market.

Participants in this study described the Vietnamese real estate landscape as "very analog" and stated that there were few PropTech applications in widespread use among professionals in the industry and even fewer that offered bilingual support to non-Vietnamese investors. Considering the increasing volume of foreign investment in the Vietnamese real estate market and the current government's support, there seems to be an opportunity for the development of PropTech that could facilitate the process of property acquisition and management for foreign investors.

While the current study did offer some insight into possible avenues of PropTech development such as property portals, social media sites, and bilingual property management software, further research must be conducted to discover the specific needs of the end users. Such research would address the limitations of this preliminary study presented here by gathering data from a larger number of participants and utilizing more robust sampling methods. By building on the evidence provided in this chapter, researchers and PropTech developers can meet the demands of the growing contingent of investors that see Vietnam as a vital property market.

**Website links reference:**

# Symbols and Acronyms

## Symbols

The proportion of variability (R2)

Composite reliability (CR)

Cronbach's (α)

## Acronyms

| | |
|---|---|
| Proptech | Property technology |
| UTAUT | Unified theory of acceptance and use of technology |
| JLL Vietnam | Jones Lang LaSalle, a real estate services company |
| SEM | Structural equation modeling |
| PLS | Partial least squares path modeling: Estimation theory |
| AVE | Average variance extracted |
| TAM | Technology Acceptance Model |
| SCT | Social Cognitive Theory |
| IDT | Innovation Diffusion Theory |
| MPCU | Model of PC Utilization complexity (social factors) |
| IDT | Innovation Diffusion Theory |
| TRA | Theory of Reasoned Action |
| DTPB | Decomposed Theory of Planned Behavior |
| TPB | Theory of Planned Behavior |
| C-TAM-TPB | Hybrid model that combines constructs of TAM and TPB |
| IoT | Internet of Things |
| OECD | Organization for Economic Co-operation and Development |
| Moen | Mobile user engagement |
| IMF | International Monetary Fund |
| GDP | Gross Domestic Product |

**Appendix - Literature summary and future studies**

| No | Author and year | Aim of the study | Findings | Recommendation/future study |
|---|---|---|---|---|
| 1 | (Siniak, 2020) | This paper reviews the situation of a phenomenon known as PropTech and focuses on its impacts on real estate industry growth. | The survey covers four finding areas: (1) PropTech applications in the real estate industry; (2) implications of PropTech for real estate market transparency; (3) how PropTech could give a region or a company a competitive advantage; and (4) concerns on the wider implications of these changes on a labor market and education. | - Dislocation of workers, and possibly unjust treatment of dissident voices by global monopolists,<br>- A proliferation of PropTech that offer new, potentially more flexible ways of working, education, matching skills, and acquiring skills.<br>- The educational base must be more flexible, universal, and – where a scientifically rigorous Real Estate course is established in the curriculum – PropTech oriented. |
| 2 | (Baum A. S., 2020) | To introduce and analyze some of the key themes of property technology (Proptech) in a global picture by implementing a | Key observations made in the report: (1) Levels of global financial investment; (2) The growth of emerging PropTech markets; | Identification of five future fields to study: (1) Smart real estate; (2) Real estate FinTech; (3) The real estate shared economy; (4) Data |
| | | deep dive analysis of Unissu's global PropTech database, the aim is to identify the stories within the stories | and (3) The emergence and 'success' of specific technology clusters. | digitalization; and (5) Smart cities. |
| 3 | (Kim Y. H., 2013) | Investigate, propose and test a Mobile user engagement (Moen) model | mobile users' engagement motivations do influence perceived value, satisfaction, and mobile engagement intention | - Design and forecast models for digital and educational environments<br>- Strategic planning for future information systems and marketing strategy research. |
| 4 | (Ullah F. S., 2019) | Present conceptual Real Estate Stakeholders Technology Acceptance Model (RESTAM) | The RESTAM framework presents a win-win potential for the key real estate stakeholders especially the two main stakeholders: end-users and agents. | - Towards accepting Big9 technologies in the real estate sector, especially for online real estate management.<br>- Online platforms such as real estate websites and cell phone applications can be investigated in detail for their technologies accepting capabilities to pave the way for technology adoption in |
| | | | | the real estate and property sector. |
| 5 | (Allameh, 2012) | Review an emerging type of dwelling, indicated as Smart Home, with a focus on future user lifestyles and needs. | the feasibility and impact of new technologies on future lifestyle and dwelling in the domain of real estate: while residential dwellings will likely become the most important hub of human life, current homes however seem to be poorly prepared for this future. | - to develop a virtual Smart Home in our future research; a prototype system<br>- to converge housing design with future needs at both macro (society) and micro (users) level |
| 6 | (Ullah F. S., 2018) | Review the adoption of disruptive (Big9) technologies in real estate. | Highlight the potential utilization of each technology for addressing consumers' needs and minimizing their regrets.<br>What, how, and who links the Big9 technologies to core consumers' needs | - to evaluate the impact of Big9 technologies in terms of how they enhance the industry according to specific stakeholder relations such as the buyer/seller interaction<br>- Based on the needs of buyers, a technology-based framework can be |
| | | | and provides a list of resources needed to ensure proper information dissemination to the stakeholders | formulated to help sellers sell their property more conveniently |
| 7 | (Ullah F. S., 2019) | to present a systems dynamics model for technology adoption in the real estate domain using critical factors and clusters of real estate websites including the system quality, information quality, service quality, and perceived ease of use. | The models work in such a way that a factor should be improved, focused on, or catered for in a real estate website. This factor will in turn add to the group directly or through its cluster, thus adding value to the group. Thus, all the presented factors, clusters, and groups if addressed properly, adds to the positive behavior of a customer for using the particular real estate service. | In the future, the model will be validated using real-life data from both customers and real estate agents and the path will be paved toward holistic TAM for real estate websites.<br><br>Further advancement in this context may involve the use of the latest 3D technologies such as laser scanning and thermal sensors for real estate technologies incorporation. |
| 8 | (Starr, 2020) | This paper explores applies Industry 4.0 to | The research is a primer on how the rapidly changing | - to experiment with emerging and maturing technologies and to adopt |
| | | commercial real estate, resulting in a framework defined here as Real Estate 4.0, a concept that encompasses FinTech and Proptech | technologies of Industry 4.0 are now disrupting and transforming real estate today into what we are calling Real Estate 4.0.<br>The future of the real estate profession and measurement of success will largely be dictated by how well we acknowledge and openly embrace these technological advances to create a built environment that is more efficient, valuable, and safer for its occupants. | agile practices that allow us to deal with uncertainty and complexity, particularly during and post the COVID-19 pandemic.<br>- While many of the new technologies are bound to be disruptive to the industry, reskilling and upskilling empower real estate professionals to become as cutting edge and successful as the technologies evolving within the profession. |
| 9 | (Arya, 2019) | This paper explores the possibility of regulating PropTech through | The authority regulating the blockchain could be provided with a majority vote on approving blocks in | to analyze the participating stakeholders and come up with a model that could recommend the most optimal distribution of |
| | | Blockchain and presents a framework for Blockchain adoption in PropTech & the role, the Government can play within this ecosystem. | the network so that all the applicable rules and regulations set are followed, and any discrepancy is reported immediately in real time. | the percentage of vote share among all the nodes, including the regulatory authority to reap the core benefits of the blockchain and also maintain the required regulations in the network |
| 10 | (Maalsen S. &., 2020) | Home, digital technologies, and data are intersecting in new ways as responses to the COVID-19 pandemic emerge. | examples show how technology is not just helping us to quarantine, access, or work and study from home but it is amplifying the porosity of the boundaries of home. | The speed at which the technology has been accepted and adopted is driven by the pandemic as people adjust their lives to lockdown but its rapidity also has implications for thoroughly thinking through their potential impacts. |
| 11 | (Ullah F. S., 2021) | Based on a mixed approach of integrated KANO-SISQual, the current study develops a comprehensive smart real estate | The KANO-based psychological assessment of the 31 factors was performed: The key attractive factors | The associated factors in the form of service, information, and system quality constructs integrate the information technology and real estate management fields. These factors can be |
| | | technology adoption model ( SRETAM) to model the users' perception of the Real estate online platforms (REOPs) in a digitally disruptive environment (DDE). | include graphical statistics, technology familiarity, well-structured content, interactive and attractive design, immersive content, self-efficacy, the novelty of information, and the presence of 3D models.<br><br>Positive perceptions about ease of use are promoted through platforms' aesthetics, quality of information, and interactivity. Thus, both the user's perceptions and platform features dictate the ease of use of REOPs. | investigated in detail in futuristic studies for individual factor impact on such platforms.<br><br>A win-win scenario can be established to enjoy more features and verified, and transparent information to make better, informed, intelligent, and regret-free rent or buy decisions. |
| 12 | (Maalsen S. &., 2019) | To observe digital technology's amplification of | By monitoring your behavior in the home, insurers, utility, and service | - The lens of digital surveillance has the potential to exacerbate |
| | | surveillance as one of the more pernicious aspects of our increasingly 'smart' homes | providers can both penalize and incentivize desired behaviors. But the COVID-19 crisis is accelerating the rate at which digital technologies cross the threshold of our homes and is re-scripting privacy in the process | these discriminatory effects.<br>- The digital discouragement of deviance and risk exercises modulatory control in the interest of profit, producing a remarkably stale and sterile smart home in the process |
| 13 | (Rondan-Cataluña, 2015) | The purpose of this paper is to provide a complete and chronological view of the evolution of the main acceptance and use of technology models, from the 1970s to the present day. | The authors have concluded that the UTAUT2 model obtains a better explanatory power than the rest of the technology acceptance models (TAMs) in the sample of mobile internet users. Furthermore, all models have a better explanation power using non-linear relationships than | To confirm models in other cultural situations this topic has been carried out with European, Asian, and Anglo-Saxon cultures. is an essential contribution. |
| | | | the traditional linear approach. | |
| 14 | (Chang, 2012) | This study reviews the most recent literature on UTAUT (Unified Theory of Acceptance, and Use of Technology) and UTAUT 2(Unified Theory of Acceptance, and Use of Technology) 2 by focusing on the findings and recommended future research. | The results confirmed that the four constructs of UTAUT contributed to Behavioral Intention even though PE seemed to be the most significant contributor.<br><br>Findings also suggest UTAUT 2 has been more explanatory and lists the suggestions for future works. The immediate implications are for researchers who wish to examine behavioral intentions and managers who wish to ensure the acceptance and use of a new system or technology. | It would also prove interesting to posit the analysis for other recent sales media such as mobile phones. |
| 15 | (Lai, 2017) | This paper contributes to the existing literature by comprehensively reviewing the concepts, applications, and development of technology adoption models and theories based on the literature review with the focus on potential application for the novelty technology of single platform E-payment. | The literature reviews share the difference between technology adoption models and theories with different theoretical insights, research problems, variables, and measurements. The development of the new theoretical research framework will depend on many factors but is not limited to the following: the research problems and objectives, gap analysis, the target market (users or developers, etc.), the organizations' goals, and the understanding of technology adoption models and theories based on the | Potential applications for technology applications for future researchers to conceptualize, distinguish and comprehend the underlying technology models and theories that may affect the previous, current, and future applications of technology adoption. |
| | | | available materials and others. | |
| 16 | (Chao, 2019) | This study developed and empirically tested a model to predict the factors affecting students' behavioral intentions toward using mobile learning (m-learning). | The results revealed that (1) behavioral intention was significantly and positively influenced by satisfaction, trust, performance expectancy, and effort expectancy; (2) perceived enjoyment, performance expectancy, and effort expectancy had positive associations with behavioral intention; (3) mobile self-efficacy had a significantly positive effect on perceived enjoyment; and (4) perceived risk had a significantly negative moderating | Future studies could employ a longitudinal design to obtain more accurate findings from a specific group. Finally, although the moderator of this study was PR, other variables such as system quality, trust, and mobile information literacy may also moderate the relationship between BI and another factor/variable. |
| | | | effect on the relationship between performance expectancy and behavioral intention. | |
| 17 | (Choi, 2020) | The research aims to Investigate factors Influencing the Behavioral Intention of Online Duty-Free Shop Users | The research confirms: (1) Attitudes were found to have the greatest influence on behavioral intentions, rather than subjective norms and desires (2) Internal influences, like family and friends, are more influenced by subjective norms than external influences such as the mass media; (3) Self-efficacy and facilitating conditions had a positive effect on perceived behavioral control. The abilities of online duty-free | Future research should be conducted on online duty-free shop users of various age groups. Moreover, since online duty-free shop users vary in nationality, in-depth studies that analyze differences between groups will be needed. Future studies will produce continuous and concrete research results that are not covered in this research. |
| | | | shop users and given facilitating conditions are helpful in behavioral control, which also affects their behavioral intention.<br><br>New finding: While existing studies have shown that the easier it is to use technology, the greater the perceived enjoyment and the more it affects attitude, this study showed different results. It indicated that if a user felt that using an online duty-free shop was easy, it was fun, but the attitude was put on hold. | |
| 18 | (Jung, 2021) | The purpose of this research is to investigate the accountability of | The results show that network externalities are essential to account | Further research needs to be conducted by adding more attributes to the sharing economy |
| | | the extended technology acceptance model (TAM) in the domain of sharing accommodation platform services. | for trust and interactivity. In addition, the results show that interactivity is an influential element in both ease of use and usefulness. Moreover, usefulness is affected by trust and interactivity. Furthermore, this research reveals the positive association between usefulness and intention to repurchase. | platform and all tourism platforms.<br>Future research considering more various methods, such as qualitative and experimental design, is needed. This could become an avenue to further determine the consumer characteristics in the sharing economy platform system. |
| 19 | (Venkatesh V. &., 2000) | The research develops and tests a theoretical extension of the Technology Acceptance Model (TAM) that explains perceived usefulness and usage intentions | Both social influence processes (subjective norm, voluntariness, and image) and cognitive instrumental processes (job relevance, output quality, result demonstrability, and | - other direct determinants of usage intentions and behavior and major contingency factors moderating the effects of perceived usefulness, perceived ease of use, subjective norm, and other constructs on the intention |
| | | in terms of social influence and cognitive instrumental processes. | perceived ease of use) significantly influenced user acceptance. | - temporal dynamics of the determinants of user acceptance.<br>- changes in work content or job goals, and changing social environments.<br>- the nature and role of social influence processes (both within teams and across teams) |
| 20 | (Mlekus, 2020) | Investigate user experience characteristics as determinants of technology acceptance. | The report's results support the extension of the technology acceptance model. The technology-inherent characteristics of output quality, perspicuity, dependability, and novelty were significant predictors of technology acceptance. | Future research should investigate further<br>technology-inherent predictors of technology acceptance. As an example, future studies could investigate if technologies that are used in the workplace and can make decisions (e.g., which task is allocated to whom) are better accepted if they are designed in adherence with organizational justice criteria. |
| 21 | (Manski, 1990) | The objective is to place an upper bound on the behavioral information contained in intentions data and to determine whether prevailing approaches to the analysis of intentions data respect the bound. | The use of intentions data to predict behavior has been controversial. The analysis of this article suggests that at least some of the controversy is rooted in the flawed premise that divergences between intentions and behavior show individuals to be poor predictors of their futures. Divergences may simply reflect the dependence of behavior on events not yet realized at the time of the survey. Divergences will occur even if responses to intentions questions are the best predictions possible | The analysis of Section 3: BOUNDS ON BEHAVIOR IMPLIED BY INTENTIONS DATA can be extended to find the best-case probability bounds implied by the responses to such questions. |
| | | | given the available information. | |
| 22 | (Fishman, 2020) | Predicting implementation: comparing validated measures of intention and assessing the role of motivation when designing behavioral interventions | The results support using an aggregate of two or three intention items that refer to the specific EBP. An even more pragmatic measure of intention consisting of a single item can also predict implementation. | In future studies of predictive validity, investigators can adapt the item stems when studying any EBP and other practitioners, such as those employed by health care organizations |
| 23 | (Warshaw, 1980) | A new intention model was developed and tested for predicting Behavioral Intentions as an alternative to the Fishbein model | Results show high stability, reliability, and productiveness across diverse test products, with very low inter-correlations between explanatory variables. | The successful application of the proposed model could provide direction for research in behavioral prediction, though refinement of questionnaire item wording is probably necessary.<br>Another research stream could focus on conceptualization and measuring the antecedents of the proposed model's five predictor variables, especially product type desire (its determinants will |
| | | | | differ for replacement purchases and new acquisitions), felt pressure from others, and the relative ability of the target brand to satisfy product use needs. |
| 24 | (Ullah F. S., 2017) | This paper systematically reviews the state-of-the-art technologies in the real estate websites of the US and Australia | Development of multiple Theoretical TAM frameworks for smart real estate management based on website design, VR & AR, Laser scanning, and 360 cameras | futuristic TAM formations for each of the disruptive technologies |
| 25 | (Munawar, 2020) | Further field-specific applications of big data in two promising and integrated fields, i.e., smart real estate and disaster management, were investigated, and a framework for field-specific applications, as well as a merger of the two areas | Further field-specific applications of big data in two promising and integrated fields, i.e., smart real estate and disaster management, were investigated, and a framework for field-specific applications, as well as a merger of the two areas through | integrations of various big data technologies and analytics tools in field-specific contexts such as data lakes and fast data.<br>most significant challenges posed by big data in specific fields such as real estate and property management or disaster management, and how technological advancements are being used to tackle these challenges. |
| | | through big data, was highlighted | big data, was highlighted | |
| 26 | (Ullah F. S.-T., 2021) | The current study assesses these DDTs adoption and innovation barriers facing the Australian real estate sector from a managerial perspective. | Technological group: the highest failure reason is attributed to T2.<br>Organizational group: the barrier with the highest failure chances for DDT adoption is the lack of organizational willingness to invest in digital marketing (O4). | System dynamics or DEMATEL techniques can be used in a future study to investigate the dynamic relationship between the barriers and the intra-relationship<br>More evenly spread respondents across all Australian states may produce different results that can be used to compare state-wise barriers for Australian real estate. |
| 27 | (Ullah F. S., 2021) | To captures the users' perception of Real Estate Online Platforms. risk and service | Risk, service, information, system, technology adoption model (RSISTAM) is proposed comprising of seven users' perceptions: risk (PR), service quality (PSEQ), information quality (PIQ), and system quality (PSYQ) | To develop a holistic framework for all-inclusive REOP adoption (including REOP web management team, real estate agents, and web developers)<br>The risks and barriers associated with the REOP technologies adoption can be studied in the future.<br>The study repeated in a developing country can be compared to the current |
| | | | from the information systems success model, and usefulness (PU), ease of use (PEU) and behavior to accept (BAU) from TAM | study to highlight the perception difference among REOP users. |
| 28 | (Felli, 2018) | This paper aims at developing and idealizing an immersive visualization mechanism for buildings using VR-based technologies to help post-construction sales. | The effects of visualizations on construction site management and processes are explored and a concept of using the data for post-construction sales is presented. | New ideas generations: using the in-construction data to enhance real estate and property sales and address post-purchase decision regrets by providing more detailed information to the customers to make informed decisions. |
| 29 | (Shirowzhan, 2019) | To implement an online tracking system for indoor positioning using radio communication systems, called the Construction Labor Tracking System (CLTS), | CLTS can monitor the approximate position of workers in detail. The developed CLTS tool is highly valuable to construction project managers because it increases the quality | The system will be used for job-site equipment and workforce monitoring during construction and will be critical for improving productivity. |
| | | and integrate it with GIS | of the collected data and enables them to measure labor productivity on construction sites. | |
| 30 | (Ullah F. S., 2019) | The study explores the Real Estate Online Platform based on information from the users' perspective in Sydney Australia | Most of the users are aware of REOPs and are using them easily. Further, the OP design and context is giving a good impression to the users. | The results are expected to lay the foundation for OP-based technology acceptance in Real Estate. |
| 31 | (Ullah F. &., 2020) | The current study aims at investigating the key post-purchase regret factors of real estate and property owners and renters over the last decade using published literature and online threads | Overall, a total of 10% and 8% increases have occurred in the regrets related to the buy-sell process and lack of inspections, respectively. On the other hand, regrets related to agents and housing costs have decreased drastically by 40% mainly due to the good return | The inspection strategies, including virtual- and augmented-realities-based virtual inspections, 360 videos, mobile laser measurements, 3D scanning, GIS-based location systems, and gadgets and drones-based inspections, are an insight into the futuristic domains of research and practice, which must be investigated in detail if the real estate sector wants to remain sustainable during |
| | | | on investments in the growing markets. | pandemic times and transform into a smart real estate sector. |
| 32 | (Low, 2020) | This study aims at understanding the principles and practices of sustainable digital marketing in the Malaysian property development industry by investigating the extent to which digital marketing has been adopted, the impediments to its adoption, and the strategies to improve digital capabilities for the local context. | The results show that the sample property development companies are driven by the benefit of easily obtaining real-time customer information for creating and communicating value to customers more effectively through the company brand. | Future research could explore other major cities of Malaysia, such as Johor Bahru or others, as well as increase the scope of the study to include developing and developed countries. |
| 33 | (Baum A. , Tokenization—The Future of Real Estate Investment?, 2021) | This article examines the mechanisms now available to tokenize real asset ownership and create active secondary markets | ▪ The author examines the mechanisms now available to tokenize real asset ownership.<br>▪ The author presents new | Further, in many land markets, fractionalization requires an intermediate structure to be established because the direct ownership of land cannot be shared amongst many, increasing the cost of |
| | | in tokenized or fractionalized units. | survey-based evidence of real estate tokenization deals.<br>▪ The author concludes that an intermediate structure is likely to be both necessary and convenient when fractionalizing a single asset. For that reason, funds are more natural targets for tokenization. | tokenization. Larger assets already held in fund structures may eventually be tokenized successfully; there may also be an alternative market for tokenized residential, social impact, or community assets where investment regulation and risk/return are not the main drivers of behavior. The mass market for the tokenization of single commercial real estate assets, however, may be some way down the road. |
| 34 | (Garcia-Teruel, 2021) | This paper proposes a model to tokenize the right of usufruct over chattels and real estate, analyzing its legal viability and limitations across six jurisdictions. | Private law rules may be adapted to the tokenization of property rights, which may contribute to the establishment of a digital market for the trading of asset-backed tokens worldwide. | Tokenization of property rights: how other limited property rights could work in blockchain or how these databases can interconnect with the public administration), |
| 35 | (Kankanhalli, 2012) | Investigations of the antecedents, mechanisms, and impacts of gamification. | - An initial review of concepts and example cases related to gamification and summarize sample applications based on their objectives, design elements, rewards, and outcomes.<br>- Articulate potential theories that can be extended to understand the motivations, design mechanisms, and impacts of gamification.<br>- Provide directions for future research in this area by outlining salient research questions on various aspects of gamification. | Based on the three themes of motivations, design, and impacts, eight potential research directions are outlined with the first three primarily related to motivations, the next two mainly related to design, and the remaining three mainly related to gamification impacts. |
| 36 | (Khobzi, 2019) | The purpose of this paper is to investigate the | The empirical findings confirm | - Future research can focus on impressions as a measure of social media marketing |
| | | impact of different ways of message framing on users' engagement behavior regarding brand posts on Facebook and to determine whether users' thumbs-up and replies moderate this impact. | that more positively and negatively framed comments result in increased user engagement. Also, an increase in the thumbs-up ratio for neutrally and negatively framed comments results in less engagement. The reply ratio might also have a positive and negative moderation effect on the influence of neutrally and positively framed comments on engagement behavior, respectively. | effectiveness on Facebook or other social media platforms.<br>- The dynamics of conversations (i.e., commenters and replies) such as the participation of new users and the return of those who already participated in the conversation, and the size of their network can identify the extent of exposure of brand posts. |
| 37 | (Gibilaro, 2020) | The paper aims to study the performance of crowdfunding REITs concerning traditional REITs to evaluate the | Results show that the performance of crowdfunding REITs is more stable over time than other REITs and the | analysis of the main features and differences among countries may allow testing if results hold independently concerning |
| | | differences in the risk-return profile and their usefulness for a diversification strategy within indirect real estate investments. | lack of correlation with traditional REITs may be exploited for constructing a more efficient diversified portfolio of indirect real estate investments. | the real estate market considered. |
| 38 | (Conway, 2018) | To define machine learning and artificial intelligence for the investor and real estate audience, examine how these new analytic, predictive, and automated technologies are being used in the real estate industry. | Machine learning and artificial intelligence can and will be used to facilitate real estate investment in myriad ways, spanning all aspects of the real estate profession -- from property management to investment decisions, to development processes -- transforming real estate into a more efficient and data-driven industry. | As the technologies develop it will be important to carefully examine their use and validity; |
| 39 | (Pham, 1997) | This paper presents a new algorithm for extracting IF-THEN rules from examples. | The algorithm employs an efficient rule searching method and a simple metric for assessing rule generality and accuracy. The paper illustrates step by step the operation of the algorithm and discusses its performance on the IRIS data classification problem. | The algorithm can be applied to other test problems, for example, recognition of design form features, mapping of manufacturing information to design features, and classification of defects in automated visual inspection. |
| 40 | (Wilson, 2018) | To study the impact when Artificial intelligence is transforming business. | While AI will radically alter how work gets done and who does it, the technology's larger impact will be in complementing and augmenting human capabilities, not replacing them. | Company roles will be redesigned around the desired outcomes of reimagined processes, and corporations will increasingly be organized around different types of skills rather than around rigid job titles. |
| 41 | (Kolbjørnsrud, 2016) | The adaptation of managers at all | A recommendation is to adopt AI to automate | The readiness of managers at all levels to assess the |
| | | levels to the world of smart machines | administration and to augment but not replace human judgment. | world with AI and automation. |
| 42 | (Chernov, 2019) | This study includes the analysis of the Artificial Intelligence usage trends and their influence on the labor market and manager's job roles | In terms of uncertainty and ambiguity, AI can't make a correct and accurate business decision based on a rational way, but in terms of complexity, it can perform well. So, in terms of business decision making the most effective method is a collaboration between managers and AI. | Intelligent machines using a rational way of decision making can offer several alternatives to the manager who has 2 options – make a decision based on his experience and intuition or let AI make a decision. |
| 43 | (Rao A. , 2017) | A Strategist's Guide to Artificial Intelligence: | As the conceptual side of computer science becomes practical and relevant to business, companies must decide what type of | The lack of capable talent — people skilled in deep learning technology and analytics — may well turn out to be the biggest obstacle for large companies. The greatest opportunities may thus be for independent businesspeople, who no |
| | | | AI role they should play. | longer need scale to compete with large companies because AI has leveled the playing field. |
| 44 | (Ngoc, 2020) | This study is intended to present opportunities and challenges for Vietnamese real estate brokerage firms in the post-Covid19 era. | In the post-Covid period, real estate brokerage firms were still able to seize many opportunities for them to return to normal. Also, many challenges when the Covid-19 pandemic is not over yet, the worrying thing about the business is its finance, and personnel, especially, the behavior of customers buying land after Covid-19, they choose safer and more risk-averse products. | To research and develop a specific legal system for the real estate mortgage market so that commercial banks have a basis to create long-term capital for real estate collateral. |
| 45 | (Clayton, 2019) | The authors discuss the changes being | The ability to adopt technology into the investment process | Proptech strategies, like other software/hardware solutions, are also likely to |
| | | brought to real estate investment by new technology and how the other articles in the special issue nest into this broad theme. | and adapt to how technology is changing the space market will be a key determinant of investment success in the future.<br><br>How closely fundamental research and new technology are related: Today's increased data availability and computing power mean that researchers can often provide new insights into even traditional real estate investment questions. | ignore national boundaries. Thus, an international presence may also be a key success factor for both the creators and the users of Proptech strategies. In today's digital, cloud-based environment, advances in data capture, management, and predictive analytics are quickly scalable to a global marketplace. |
| 46 | (Porter, 2019) | Examine the co-emergence of PropTech and Big Data, the impact on land and housing dynamics, and the implications for | Digital technologies utilizing big data and artificial intelligence will remain on the sidelines of planning practice. As | to proactively formulate how technology should fit within the planning system, how the planning system fits within smart cities, and how their processes might |
| | | planning governance and systems | technologies mature and become more sophisticated their utility will be exploited to aid design and assessment. | change to accommodate these shifts. |
| 47 | (Sittler, 2017) | the report aims to analyze future business models and trends in real estate. | With (Nearly) Complete PropTech List with each business model,<br>► Digitalization has a growing importance in the real estate industry<br>► PropTech sector is at the beginning | ► Innovation, flexibility, and communication are needed<br>► Ongoing changing in society and customer behavior<br>► New business models are occurring |
| 48 | (Delclós, 2020) | Housing in the digital age: Trends and implications | the implications of digitalization are systemic, with consequences that cut across a variety of social fields. | challenges for urban housing policies in particular, including telework-enabled urban flight, rising residential segregation, and a growing digital divide. |
| 49 | (Putatunda, 2019) | This paper proposes a machine learning | The Random Forest method is the best performer in terms | The proposed methods can be used for other applications of PropTech. |
| | | approach for solving the house price prediction problem in classified advertisements. | of prediction accuracy. | The error margins can be reduced further if we use much larger datasets, which we aim to work on in the future. |
| 50 | (Argelich Comelles, 2020) | Various proposals for the regulation of peer-to-peer accommodation, considering the required Covid-19 social distancing as an opportunity | Spanish | Spanish |
| 51 | (Di Giorgio, 2020) | To analyze in detail the change currently underway in the sector, starting from an analysis of the key aspects and then reaching an illustration of the main models and | Numerous obstacles and threats to the adoption of Proptech in the real estate sector, including operational, regulatory, and social obstacles. | numerous advantages both to market operators and the community in general, but to truly exploit its benefits all actors must understand the innovation and adapt accordingly |
| | | actors present in the context. | | |
| 52 | (Deloitte, 2021) | Real Estate Predictions 2020: to analyze the latest developments and trends that will impact business. | Identify the impact of climate change, Conversational AI, Proptech, data minimization, working place to living environments, and smart development of smart places. | There's growing interest among enterprises in looking beyond what's new to what's next. |
| 53 | (Gupta, 2021) | Impact of Covid-19 pandemic to house price and rent. | The COVID-19 pandemic brought house price and rent declines in city centers, and prices and rent increase away from the center, thereby flattening the bid-rent curve in most U.S. metropolitan areas.<br>Across MSAs, the flattening of the bid-rent curve is larger when working from | Housing markets predict that urban rent growth will exceed suburban rent growth for the foreseeable future. |
| | | | home is more prevalent, housing markets are more regulated, and supply is less elastic. | |
| 54 | (KPMG, 2019) | An annual review of the real estate industry's journey into the digital age: who has responsibility for advancing the digital agenda and whether companies have the talent in place to deliver the job. | Property companies are increasing their engagement with digital, and many companies are taking digital developments 1in their stride. Brokers and advisors, in particular, are at the forefront of the technological revolution, with automation and digitalization playing a growing part in their day-to-day operations. | Like the other aspects of digital transformation, cyber security provides an opportunity for forward-thinking companies to differentiate themselves from their peers. |
| 55 | (Sing, 2021) | Examines the effects of online listing portals on success rates in sales and | (1) the use of PropTech increases the success rate in | The writing style analysis in this paper exploits the methodologies of textual analysis, which combines knowledge from statistics |
| | | performance of real estate agents in the private housing markets in Singapore | converting listings into sales;<br><br>(2) conditional on the success of the sales, real estate agents sell houses at an average premium of 3.26% when they adopt PropTech relative to sales using the conventional approach; and<br><br>(3) the adoption of PropTech significantly enhances real estate agents' performance in terms of a higher price gap and shorter time on the market via the effort channel, the information channel, and the portal selection. | and linguistics and is sometimes referred to as computational linguistics, natural language processing, or information retrieval. Initially developed as a means to automate the understanding of textual data with machines, a task considered too slow or too complex for humans to take on, textual analysis has since been extended to model and structure information from a variety of textual sources. |
| 56 | (Friedman, 2020) | Draw insights about PropTech by taking a | Images of building facades hold some information that can | Doing a higher-fidelity analysis of facades to pinpoint exactly what in |
| | | decidedly different approach; one which uses technical experimentation, data analysis, and industry immersion as a basis for examining the broader industry from the inside | be used to value a building.<br>Context-specific notions of value probed the conditions which produce a such large geospatial variance in the automated valuation model (AVM) to break down problems to drive insights, showing the potential for this technology to be used to highlight issues beyond property value. | the facade contains information that can be used for valuation.<br>Future work might apply this AVM to understand how the values of building elements change over time as a way to measure the cultural perception of value.<br>Design technical experiments around probing very specific relationships between industry, technical matters, and externalities in PropTech |
| 57 | (Andreasson, 2019) | Examine how digital solutions and services can contribute to value creation in the Swedish real estate sector. | Digitalization allows real estate to be used in new innovative ways to create value for both tenants and real estate companies. | - economic aspects, laws and regulations, IT security, or personal integrity on Proptech<br>- infrastructure (hardware and software solutions. digital solutions and services) to implement |
| | | | | digitalization in the real estate sector.<br>- business models, suitable for innovative work concerning digitalization in the Swedish real estate sector. |
| 58 | (RICS Research Trust, 2020) | Critically review the current and potential applications of distributed ledger technology (DLT), particularly in the form of blockchain in real estate, and discuss how stakeholders in the industry will be affected. | • Blockchain-based systems are not completely trustless, with real estate transactions, intermediaries are still needed to verify sources of data in transaction value chains.<br>• Efficiency and scalability remain major concerns for the industry and end-users.<br>• Legally, existing regulations and laws are still catching up with technological developments, and | • Stakeholders should work together (Deloitte, 2017b; Baum, 2017), as a single group is unlikely to be able to drive the changes alone (Saull and Baum,2019).<br>• Need for governments to evaluate the impact of the technology and provide up-to-date guidance.<br>• The report focuses on blockchain, as the applications of distributed ledger technology (DLT) are currently dominated by blockchain. Other types of DLT should be explored. For example, |
| | | | there are many uncertainties in the self-executing nature of smart contracts and the ownership and rights of tokenized assets.<br>• Socially, the prerequisite for the technology to be applied successfully in the real estate sector is the participation and coordination of all stakeholders involved. Without guidance from the government and regulators, such coordination is difficult to achieve. Technology is still misunderstood. | directed acyclic graph (DAG) and Holo-chain are being explored in terms of increasing efficiency and scalability. |
| 59 | (Savills, 2019) | Savills global real estate markets report 2009 which | The relationship between the physical and virtual worlds of | Future design and development have to be as flexible and as mixed-use as possible” |
| | | focuses on disruption. | competitive gaming is creating the need for a whole new model of real estate | |
| 60 | (Braesemann, 2020) | To investigate whether PropTech is turning real estate into a data-driven market. | PropTech is indeed an increasingly important, global phenomenon, with data analytics technologies at the core of the network of property technologies | Users and owners of real estate, to benefit from the efficiency gains associated with the digitalization of the market, need to become aware of the business value of data they are generating in buying, renting, or managing real estate. |
| 61 | (Feth, 2018) | To study the role of PropTech in the changing Real Estate Industry. | This article illustrates the different segments that digital solution providers are split into and explains their importance for the real estate sector | Barriers to the entrance to the Proptech industry |
| 62 | (Kempeneer, 2021) | Identify key gaps in the literature and practice and provides a framework to | Transition to user-centered smart real estate is the solution to improving both the environmental | Critically evaluate and contextualize the ESG framework they are using as well as the extent to which users are |
| | | further the understanding of how environment, social, and governance (ESG) factors can add societal and financial value to the real estate sector | (E) and social (S) sustainability of buildings, as well as their investment value | considered and smart technology is employed. |
| 63 | (Wang, 2021) | To explore the NFT ecosystem | - provide technical components, protocols, standards, and desired proprieties.<br>- security evolution, with discussions on the perspectives of their design models, opportunities, and challenges | Opportunities (Boosting gaming industry, Flourishing virtual events, Protecting digital collectibles, Inspiring the Metaverse)<br>Challenges (Usability: slow confirmation, High gas prices; Security and Privacy Issues; Governance Consideration; Extensibility issues) |
| 64 | (Bamakan, 2021) | To examine the requirements of presenting intellectual property assets, | Offering a layered conceptual NFT-based patent framework with a comprehensive discussion on each | Future studies are suggested to upgrade the consensus method used in the verification layer. |
| | | specifically patents, as NFTs. | layer, including storage, decentralized authentication, decentralized verification, Blockchain, and application layer | Different NFT standards pose cross-platform problems due to centralized marketplaces like Raible and OpenSea. |
| 65 | (Kim J. &., 2021) | An Integrated Analysis of Value-Based Adoption Model and Information Systems Success Model for PropTech Service Platform | the consumer's intention to use services is influenced by service practicability in terms of consumer value. The intention to continue using a service is influenced by user satisfaction. | More detailed analysis according to product classification. This business ecosystem aims to provide optimized product information using big data obtained through the combination of the latest information technologies. This aspect can also motivate future research. Future studies can also focus on the optimization service of AI based on value analysis or the provision of non-face-to-face services through virtual or augmented reality. |
| 66 | (Venkatesh V. M., 2003) | To (1) review user acceptance literature and discuss eight prominent models, (2) empirically compare the eight | UTAUT thus provides a useful tool for managers needing to assess the likelihood of success for new technology introductions and | Developing a deeper understanding of the dynamic influences studied here, refining measurement of the core constructs used in |
| | | models and their extensions, (3) formulate a unified model that integrates elements across the eight models, and (4) empirically validate the unified model. | helps them understand the drivers of acceptance to proactively design interventions (including training, marketing, etc.) targeted at populations of users that may be less inclined to adopt and use new systems. | UTAUT, and understanding the organizational outcomes associated with new technology use. |
| 67 | (Rahi S. G., 2018) | Investigating the role of a unified theory of acceptance and use of technology (UTAUT) in the internet banking adoption context | The results have indicated support for UTAUT findings by Venkatesh et al. (2003). Performance expectancy was significant with users' intention to adopt internet banking suggesting that users having more performance expectancy had more inclined towards adoption of internet banking. Interestingly, the effect of sizes revealed that social influence, facilitating condition, and effort expectancy had small | Future research can extend the UTAUT model with other variables such as website design, assurance, reliability, and e-customer service to get more in-depth knowledge about factors that propel- ling user's intention to adopt internet banking. |
| | | | effect sizes in predicting users' intention to adopt internet banking. | |

**Declaration of Authorship**

I, LE TUNG Bach, hereby declare that this dissertation/thesis entitled: “The behavioral intention to adopt proptech services in Vietnam real estate market” and the work presented in it are my own and has been generated by me as the result of my own original research. I declare that I have authored this thesis independently, that I have not used other than the declared sources/resources, and that I have explicitly marked all material which has been quoted either literally or by content from the used sources. According to my knowledge, the content or parts of this thesis have not been presented to any other examination authority and have not been published. Where any part of this dissertation has previously been submitted for a degree or any other qualification at this university or any other institution, this has been clearly stated. Where I have used or consulted the published work of others, this is always clearly attributed. Where I have quoted from the works of others, the source is always given. With the exception of such quotations, this dissertation is entirely my own work.

Hanoi, Date: November 30, 2022

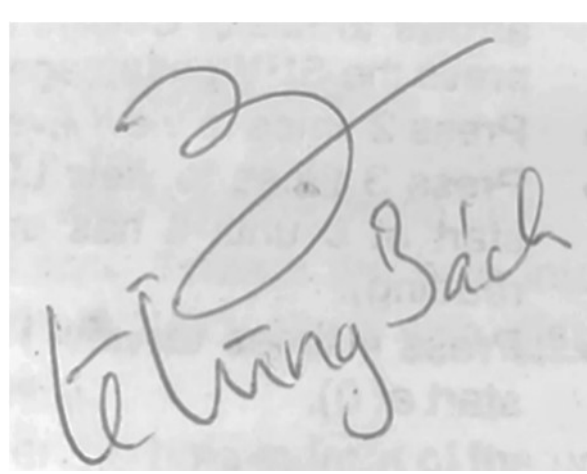

×